\documentstyle[aps,prb,psfig,epsf,eqsecnum]{revtex}

\tighten

\begin{document}

\draft

\title{Between Poisson and GUE statistics: Role of the Breit--Wigner width}

\author{Klaus M. Frahm$^1$, Thomas Guhr$^2$ 
        and Axel M\"uller--Groeling$^2$}
        
\address{$^1$Laboratoire de Physique Quantique, UMR 5626 IRSAMC, 
Universit\'e Paul Sabatier, 
F--31062 Toulouse, France \\
$^2$Max--Planck--Institut f\"ur Kernphysik, Postfach 103980,
                       D--69029 Heidelberg, Germany}
 
\date{\today}
 
\maketitle

\begin{abstract}
  We consider the spectral statistics of the superposition of a random
  diagonal matrix and a GUE matrix. By means of two alternative
  superanalytic approaches, the coset method and the graded eigenvalue
  method, we derive the two--level correlation function $X_2(r)$ and
  the number variance $\Sigma^2(r)$. The graded eigenvalue approach
  leads to an expression for $X_2(r)$ which is valid for all values of
  the parameter $\lambda$ governing the strength of the GUE admixture
  on the unfolded scale. A new twofold integration representation is
  found which can be easily evaluated numerically. For $\lambda \gg 1$
  the Breit--Wigner width $\Gamma_1$ measured in units of the mean
  level spacing $D$ is much larger than unity. In this limit, closed
  analytical expression for $X_2(r)$ and $\Sigma^2(r)$ can be derived
  by (i) evaluating the double integral perturbatively or (ii) an {\it
    ab initio} perturbative calculation employing the coset method.
  The instructive comparison between both approaches reveals that
  random fluctuations of $\Gamma_1$ manifest themselves in
  modifications of the spectral statistics. The energy scale which determines
  the deviation of the statistical properties from GUE behavior is
  given by $\sqrt{\Gamma_1}$. This is rigorously shown and discussed 
  in great detail. The Breit--Wigner $\Gamma_1$ width
  itself governs the approach to the Poisson limit for $r\to\infty$.
  Our analytical findings are confirmed by numerical simulations of an
  ensemble of $500\times 500$ matrices, which demonstrate the
  universal validity of our results after proper unfolding.
\end{abstract}


\section{Introduction}
\label{sec1}

One of the archetypical problems in quantum mechanics consists of
calculating (certain properties of) the eigenvalue spectrum of a
diagonal operator and a superimposed non--diagonal one. Little can be
said in general about this problem. In our paper, we focus on the
particular case where the matrix representations $H_0$ and $H_1$ of
the above operators can be taken from the Poisson and the Gaussian
Unitary Ensemble~\cite{Meh91} (GUE), respectively,
\begin{equation}
H = H_0 + \alpha H_1 \ ,
\label{eq1}
\end{equation}
with $\alpha$ some strength parameter. The Poisson Ensemble is
constructed from all those matrices whose eigenvalues are independent
random numbers with identical, and largely arbitrary, distribution
function. Due to the rotational invariance of the GUE, it suffices to
consider only diagonal matrices $H_0$ of the Poisson type. The random
matrix model (\ref{eq1}) should provide an adequate description for
numerous transition phenomena from regular to chaotic fluctuation
properties in atomic, nuclear, and condensed matter physics, as well
as in quantum chaology (for a review see Ref.~\onlinecite{Guh97a}).

Nuclear physics provides an important example. In heavy ion reactions,
fast rotating compound nuclei are produced. The rotation is a
collective motion of all nucleons which is often accompanied by single
particle excitations. Thus the total Hamiltonian can be modeled as a
sum of two contributions, a regular one, $H_0$, describing the
collective motion, and a stochastic one, $H_1$, describing the
influence of the single particle excitations. This scenario was
studied numerically in Ref.~\onlinecite{Guh89} for the cases that
$H_0$ is drawn from a Poisson or a harmonic oscillator ensemble. Using
Efetov's supersymmetry method~\cite{Efe83,Ver85}, the qualitative
behavior of the two--level correlations was also discussed
analytically in Ref.~\onlinecite{Guh89}.

As a further important example we mention the problem of two
interacting particles in a random potential introduced by
Shepelyansky~\cite{She94}. He predicted that two particles in a
one--dimensional disordered chain can be extended on a scale $L_2$ far
exceeding the one--particle localization length $L_1$.  Subsequent
work~\cite{Imr95,Fra95,Opp96} quickly led to a definite confirmation
and better understanding of this phenomenon. One possible
approach~\cite{She94} to this problem is to construct an effective
Hamiltonian by diagonalizing the noninteracting part of the
two--electron problem and expressing the microscopic Hamiltonian in
the basis of two--electron product states. The resulting
representation consists of a diagonal contribution containing the
eigenvalues of the noninteracting problem, and an off--diagonal
contribution originating from the interaction operator. With the
crucial assumption that both the diagonal and the off--diagonal matrix
elements can be chosen to be random variables we arrive at the above
random matrix model (for system sizes $L$ of the order of $L_1$).  In
the regime $L > L_1$ the effective Hamiltonian has been studied in
some detail~\cite{Jac95,Mir95,Fra95a}.  Imry has shown~\cite{Imr95}
that the enhancement factor $L_2/L_1$ is given by the ``Breit--Wigner
width'' or ``two--particle Thouless energy'' $\Gamma_2$ measured in
units of the two--particle level spacing $D_2$, $L_2/L_1 = \Gamma_2$.
This raised, among other things, the question how to identify
$\Gamma_2$ in typical spectral observables like the number variance
$\Sigma^2(r)$ of the random matrix model (\ref{eq1}) considered
here~\cite{Wei96}. This problem, and some surprises which we
encountered while studying it, have been our original motivation in
this project.

The random matrix model (\ref{eq1}) has been considered by many
authors~\cite{Fre88,Ley90,Len92,Pan95,Guh96,Guh97,Alt97}. Recently,
one of us succeeded~\cite{Guh96} in deriving exact integral
representations for the spectral $k$--point functions of the
Hamiltonian (\ref{eq1}).  To this end, the graded eigenvalue method
was used~\cite{Guh96,Guh91}, which is a variant of Efetov's
supersymmetry method~\cite{Efe83,Ver85}.  In an effort to discuss
certain approximations from these integral representations, two of the
present authors derived~\cite{Guh97} a closed expression for the
spectral two--point function $X_2(r)$ by means of a special kind of
saddle--point approximation. This led to the surprising observation
that the energy scale at which $X_2(r)$ and the number variance
$\Sigma^2(r)$ deviate from random matrix behavior is given by
$\sqrt{\Gamma_1}$ and not by $\Gamma_1$. Here, $\Gamma_1$ is the
Breit--Wigner (or spreading) width induced by the perturbation $\alpha
H_1$ and corresponds to the quantity $\Gamma_2$ in the above example.
Both $\Gamma_1$ and $r$ are measured in units of the mean level
spacing $D$. This result was later confirmed in a perturbative
calculation~\cite{Alt97} of the two--point correlator $X_2(r)$. The
saddle--point approximation employed in Ref.~\onlinecite{Guh97} was
limited to certain situations and does not include cases in which the
energy separations $r$ is much larger than the Breit--Wigner width
$\Gamma_1$.

The purpose of the present paper is threefold. First, we derive,
avoiding the above--mentioned saddle--point expression, a relatively
simple exact expression for $X_2(r)$ for all $r$ and all relative
strengths of $H_0$ and $H_1$. For $\Gamma_1 \gg 1$ this leads to very
compact analytical formulas for both $X_2(r)$ and $\Sigma^2(r)$. The
two scales $\sqrt{\Gamma_1}$ and $\Gamma_1$ can be identified and
interpreted. Second, we compare both variants of the supersymmetry
formalism, the graded eigenvalue method and the coset method, by
deriving our results independently for both methods. This sheds some
light on the relation between these approaches. Furthermore our
comparison elucidates the role played by statistical fluctuations of
the Breit--Wigner width. Third, we confirm our result with the help of
rather extensive numerical simulations of the spectral properties of
the Hamiltonian (\ref{eq1}).

Our paper is organized accordingly. Following this introduction,
Sec.~\ref{sec2} deals with the graded eigenvalue method and the
improved treatment of the integral representations derived in
Ref.~\onlinecite{Guh96}. In Sec.~\ref{sec3} the coset method is
invoked to essentially re-derive the results of Sec.~\ref{sec2}.  We
have tried to keep both sections reasonably self--contained. Readers
who are completely unfamiliar with the supersymmetry formalism should,
however, consult the introductory literature for the
coset~\cite{Efe83,Ver85} and the graded eigenvalue
method~\cite{Guh91,Guh96}. Section~\ref{sec4} is devoted to our numerical
simulations and Sec.~\ref{sec5} contains a summary and the discussion.

\section{Graded eigenvalue method}
\label{sec2}

In a first part, Sec.~\ref{sec2_1}, we introduce some basic
terminology and briefly recapitulate the derivation of the integral
representations~\cite{Guh96} of the $k$--level correlation functions for
the convenience of the reader. In Sec.~\ref{sec2_2} the integral
representation for the two--point function is further transformed to
derive a relatively simple and, most importantly, tractable expression
for the level--level correlator. Finally, Sec.~\ref{sec2_3} is devoted
to a perturbative evaluation of this expression in the limit where the
Breit--Wigner width is much larger than the level spacing.

\subsection{Integral representation for spectral correlation
  functions}
\label{sec2_1}

The $k$--level correlation functions for the Hamiltonian $H$ in
Eq.~(\ref{eq1}) characterize the spectral statistics completely and
are defined by
\begin{equation}
R_k(x_1,\ldots,x_k,\alpha) = \frac{1}{\pi^k} \int
     P_N(H) \prod_{p=1}^k {\rm Im } \, {\rm tr} \frac{1}{x_p^- - H} d[H] \ .
\label{eq2}
\end{equation}
The probability of finding $k$ energy eigenvalues in infinitesimal
intervals $dx_i$ around $x_i$ ($i=1,\ldots,k$) is given by
$R_k(x_1,\ldots,x_k) dx_1 \ldots dx_k$. For the case considered in
this paper the probability distribution function $P_N(H)$, which
depends explicitly on the dimension $N$ of the matrices $H$, is 
given by the product
\begin{equation}
P_N(H) = P_N^{(0)}(H_0) P_N^{(1)}(H_1)
\label{eq3}
\end{equation}
with
\begin{eqnarray}
P_N^{(0)}(H_0) &=& \prod_{n=1}^N p^{(0)}(H_0^{nn}) \prod_{n>m}
\delta({\rm Re} H_0^{nm}) \delta({\rm Im}H_0^{nm}) \ , \nonumber\\
P_N^{(1)}(H_1) &=& \frac{2^{N(N-1)/2}}{\pi^{N^2/2}} 
 \exp\left( -{\rm tr}(H_1^2) \right) \ .
\label{eq4}
\end{eqnarray}
The function $p^{(0)}(H_0^{nn})$ is smooth but otherwise arbitrary.
We introduce the modified $k$--point correlators 
$\hat{R}_k(x_1,\ldots,x_k)$, which are obtained by omitting the
projection onto the imaginary part in Eq.~(\ref{eq2}). The original
quantities can be reconstructed by appropriate linear combinations of
the $\hat{R}_k$. To perform the ensemble average we write the modified
correlators in terms of a supersymmetric normalized generating
functional, 
\begin{equation}
\hat{R}_k(x_1,\ldots,x_k,\alpha) = \frac{1}{(2\pi)^k}
\frac{\partial^k}
{\partial J_1\ldots \partial J_k}
Z_k(x+J,\alpha)\Bigg\vert_{J=0} 
\label{eq5}
\end{equation}
where the energies and source variables form diagonal $2k\times 2k$
matrices according to $x={\rm diag}(x_1,x_1,\ldots,x_k,x_k)$ and
$J={\rm diag}(-J_1,J_1,\ldots,-J_k,J_k)$, respectively.
The averaged functional takes the form
\begin{equation}
Z_k(x+J,\alpha) = \int d[H_0] P_N^{(0)}(H_0)
                  \int d[\sigma] Q_k(\sigma,\alpha) {\rm detg}^{-1}
                  [(x^\pm+J-\sigma)\otimes 1_N - 1_{2k}\otimes H_0] \ ,
\label{eq6}
\end{equation}
where $\sigma$ is a $2k\times 2k$ Hermitean supermatrix and
$Q_k(\sigma,\alpha) = 2^{k(k-1)}\exp(-{\rm trg}\sigma^2/\alpha^2)$ is
a normalized graded probability density. The subsequent steps can be
summarized as follows~\cite{Guh91}. The matrix $x+J$ is shifted from
the graded determinant to the graded probability density and the
supermatrix $\sigma$ is diagonalized according to $\sigma=u^{-1}su$,
where $s={\rm diag}(s_{11},is_{12},\ldots,s_{k1},is_{k2})$. The volume
element can be rewritten as $d[\sigma]=B_k^2(s) d[s] d\mu(u)$ with
$B_k(s)$ the Jacobian (called {\it Berezinian} in this case) of the
transformation.  The non--trivial integration over the unitary
diagonalizing supergroup with its Haar measure $d\mu(u)$ is the
central step in the graded eigenvalue method and can be performed with
the supersymmetric extension of the Harish--Chandra Itzykson Zuber
integral. Collecting everything we arrive at
\begin{eqnarray}
Z_k(x+J,\alpha) &=& 1-\eta(x+J) \, + \, \frac{1}{B_k(x+J)} 
                    \int G_k(s-x-J,\alpha) Z_k^{(0)}(s) B_k(s)
                    d[s] \ , \nonumber\\
Z_k^{(0)}(x+J) &=& \int d[H_0] P_N^{(0)}(H_0) {\rm detg}^{-1}
                 [(x^\pm+J)\otimes 1_N - 1_{2k}\otimes H_0] \ ,
\label{eq7}
\end{eqnarray}
where the kernel resulting from the group integration is Gaussian and
given by
\begin{equation}
G_k(s-x,\alpha) = \frac{1}{(\pi\alpha^2)^k}
          \exp\left(-\frac{1}{\alpha^2}{\rm trg}(s-x)^2\right) \ .
\label{eq7a}
\end{equation}
The distribution $1-\eta(x+J)$ in Eq.~(\ref{eq7}) ensures the
normalization $Z_k(x,\alpha)=1$ at $J=0$. We mention in passing that
the generating function $Z_k(x+J,\alpha)$ satisfies an {\it exact}
diffusion equation in the curved space of the eigenvalues of Hermitean
supermatrices. Here, $t=\alpha^2/2$ is the diffusion time and the
generating function $Z_k^{(0)}(x+J)$ serves as the initial condition.
This diffusion is the supersymmetric analogue~\cite{Guh96,Guh97b} for
the diffusion of the probability distribution function in the space of
ordinary matrices, which is equivalent to Dyson's Brownian. 
Performing the source term derivatives in Eq.~(\ref{eq5}) we find
\begin{equation}
R_k(x_1,\ldots,x_k,\alpha) = \frac{(-1)^k}{\pi^k}
  \int G_k(s-x,\alpha) \Im Z_k^{(0)}(s) B_k(s) d[s] \ , 
\label{eq8}
\end{equation}
where the symbol $\Im$ indicates the above--mentioned proper
linear combination of terms. In order to arrive at truly universal
quantities we have to measure all energies in units of the mean level
spacing $D$. This leads to the definitions $\xi_p=x_p/D$, $\lambda =
\alpha/D$, $X_k(\xi_1,\ldots,\xi_k,\lambda) =
\lim_{N\to\infty}D^kR_k(x_1,\ldots,x_k,\alpha)$, and $z_k^{(0)}
=\lim_{N\to\infty} Z_k^{(0)}(Ds)$. The level correlation functions
$X_k$ defined in this way are translation invariant over the spectrum.
They can be expressed as
\begin{equation}
X_k(\xi_1,\ldots,\xi_k,\lambda) = \frac{(-1)^k}{\pi^k}
\int G_k(s-\xi,\lambda) \Im z_k^{(0)}(s) B_k(s) d[s]
\label{eq9}
\end{equation}
after the redefinition $s\to s/D$. We notice the very similar
structure of the integral representations~(\ref{eq8}) and~(\ref{eq9}).
This is so because the above mentioned diffusion in superspace is,
apart from the initial conditions, the same on all scales, in contrast
to the diffusion in ordinary space which is equivalent to Dyson's
Brownian.  The two--point function $X_2$, to which we restrict our
attention henceforth, depends on $r=\xi_1-\xi_2$ only and we can
perform two of the four integrations in Eq.~(\ref{eq9}) due to
translational invariance. This leads to the integral
representation~\cite{Guh96}
\begin{equation}
X_2(r,\lambda) = -\frac{8}{\pi^3\lambda^2}
\int\limits_{-\infty}^\infty \int\limits_{-\infty}^\infty
     \exp\left( -\frac{1}{2\lambda^2}(t_1^2+t_2^2)\right)
     \sinh\frac{rt_1}{\lambda^2} \sin\frac{rt_2}{\lambda^2}
     \frac{t_1t_2}{(t_1^2+t_2^2)^2} \Im z_2^{(0)}(t_1,t_2) dt_1dt_2
\label{eq10}
\end{equation}
with $t_1 = s_{11}-s_{21}$ and $t_2=s_{12}-s_{22}$. The initial
condition $z_2^{(0)}$ is still arbitrary.
For the case of Poisson regularity we have
\begin{equation}
\Im z_2^{(0)}(t_1,t_2) = \frac{1}{2}{\rm Re}\left(
   \exp\left( -i\pi\frac{t_1^2+t_2^2}{2t_1^-} - 1 \right)\right) \ .
\label{eq10aa}
\end{equation}
Equation~(\ref{eq10}) is exact but, unfortunately, difficult to
evaluate as it stands. In the following subsection, we derive a more
convenient formulation of Eq.~(\ref{eq10}).

\subsection{Simplification of the two--point level correlation
  function $X_2(r,\lambda)$} 
\label{sec2_2}

\begin{figure}
\centerline{
\psfig{file=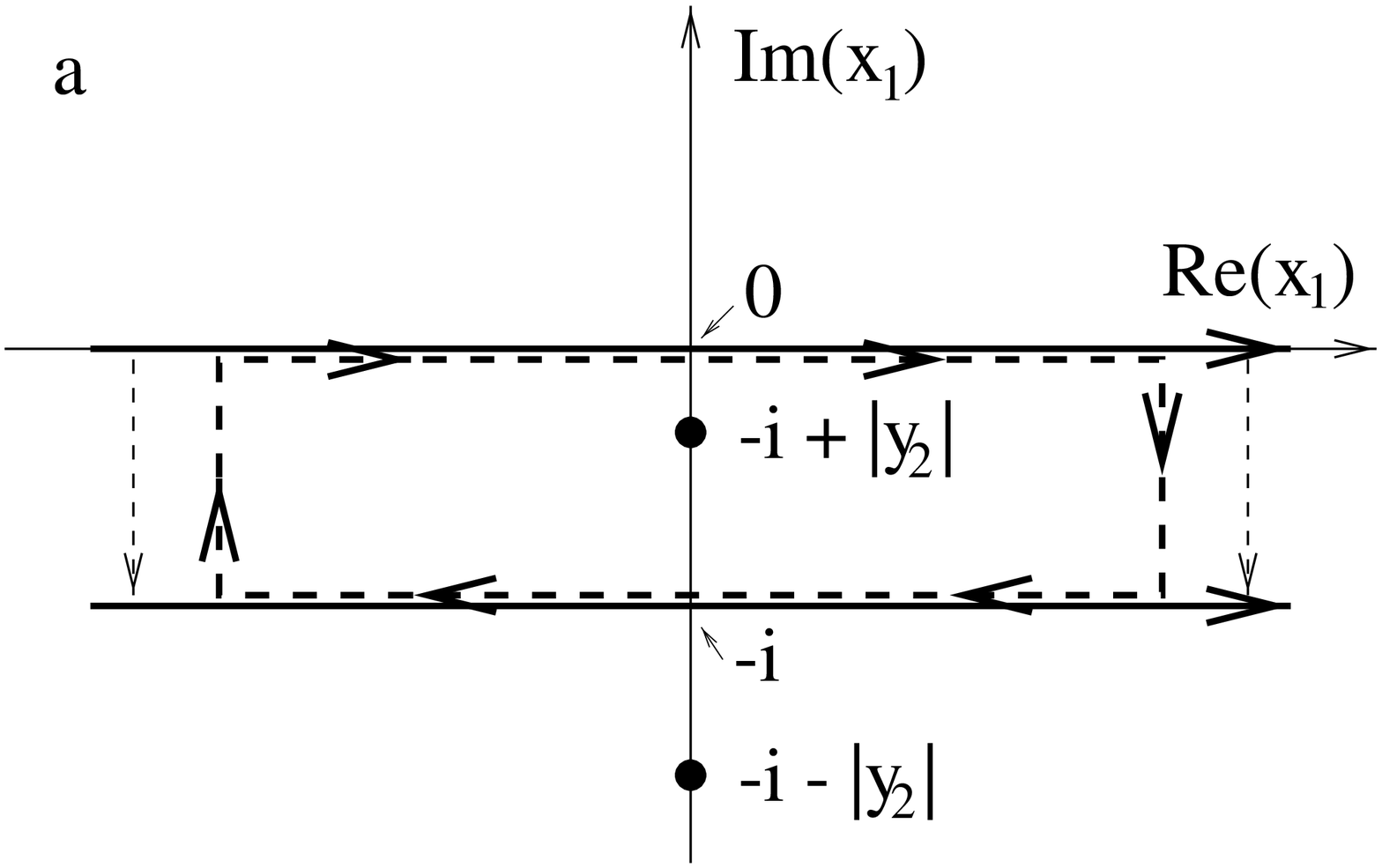,width=3.0in}
}
\vspace{0.5cm}

\centerline{
\psfig{file=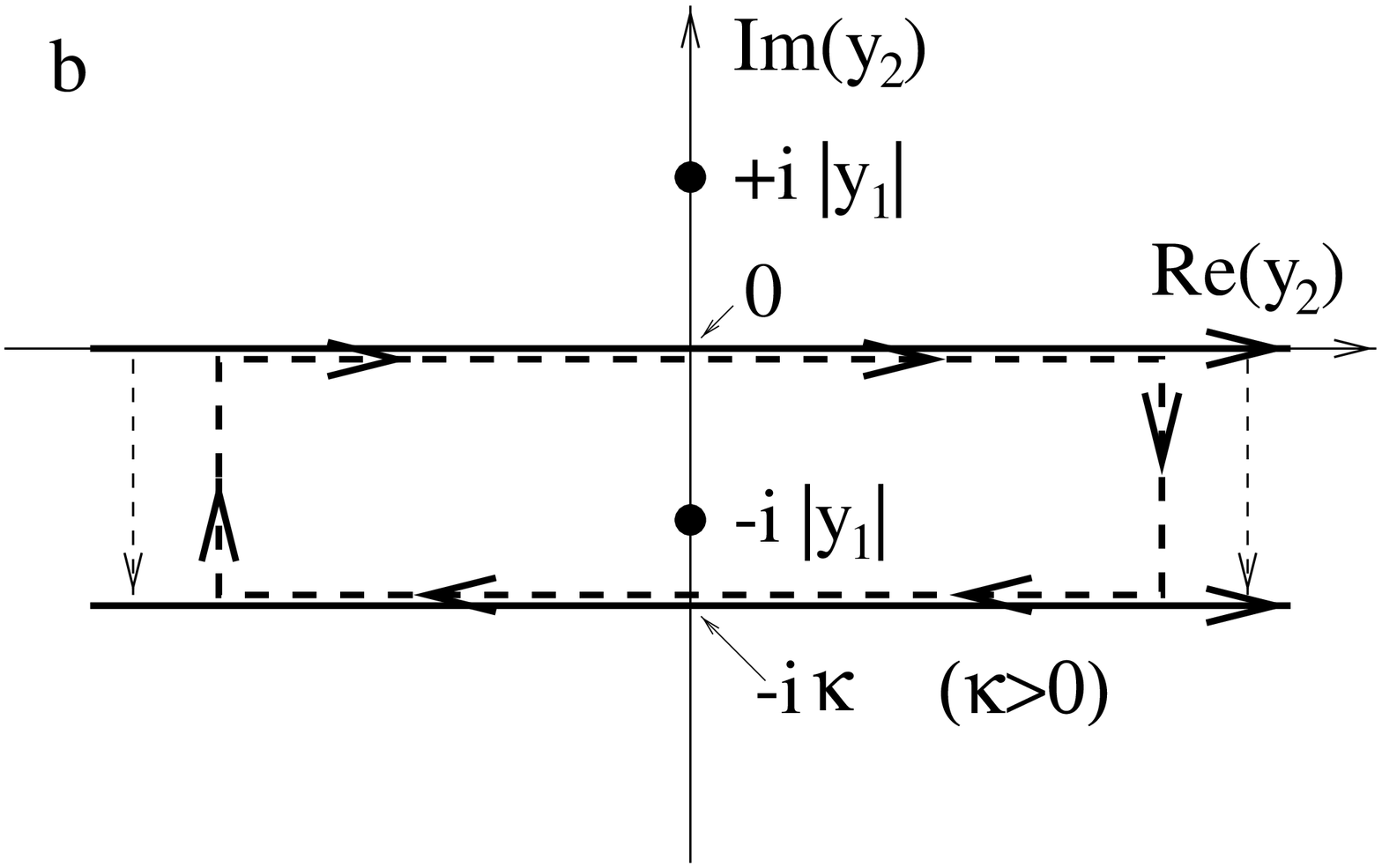,width=3.0in}
}
\vspace{0.5cm}

\centerline{
\psfig{file=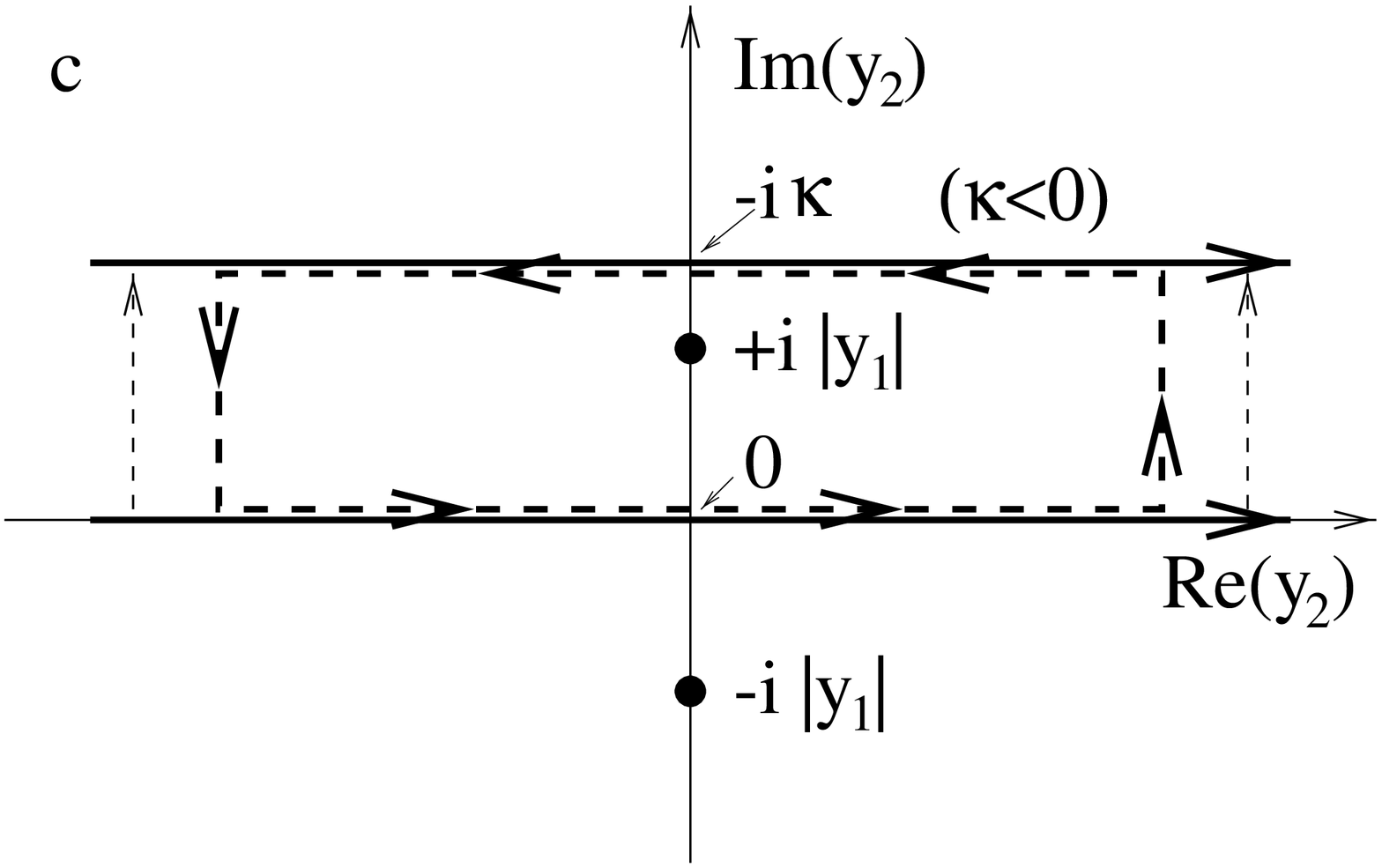,width=3.0in}
}
\vspace{0.5cm}

\caption{Graphical illustration of the shifts of the complex integration
paths. The two integration paths (full lines) differ by the closed contour 
integral (dashed line). The direction of the shifts are indicated 
by the two thin--dashed lines. The first shift $y_1=x_1+i$ is shown in (a).
Here for $|y_2|<1$ the pole $-i+|y_2|$ lies inside the contour
and results in the residuum contribution $R_1$ [see Eq.~(\protect\ref{eq15})].
The second shift $u_2=y_2+i\kappa$ is illustrated in (b) (for $\kappa>0$)
and (c) (for $\kappa<0$). The singularity $\pm i|y_1|$ lies inside the
contour if $|y_1|<|\kappa|$. The corresponding residuum contribution
is given by $R_2$ in Eq.~(\protect\ref{eq15}). 
}
\label{fig1}
\end{figure}

As explained in the introduction, a previous effort~\cite{Guh97} to
extract physical information from Eq.~(\ref{eq10}) led to the
discovery that the energy scale at which $X_2(r,\lambda)$ deviates
from random matrix behavior is linear in $\lambda$. For very general
reasons, however, the Breit--Wigner width is always quadratic in the
strength of the perturbing matrix elements, hence
$\Gamma_1\propto\lambda^2$. This led to the conclusion that
$\sqrt{\Gamma_1}$ is the important energy scale.  The approximation
used to derive this result was, however, not valid for $r \gg
\Gamma_1$. Here, we avoid the saddle--point approximation proposed in
Ref.~\onlinecite{Guh97} and proceed as follows.  We introduce the new
variables $x_j = t_j/(\pi\lambda^2)$ and the important abbreviations
\begin{equation}
c=\frac{1}{\pi^2\lambda^2}\ , \quad\quad \kappa =
\frac{r}{\pi\lambda^2} \ .
\label{eq10a}
\end{equation}
It is instructive to note that
$\pi\lambda^2$ is actually the Breit--Wigner width measured in units
of the level spacing, $\pi\lambda^2 = \Gamma_1$. This can be deduced
from the local density of states for the model studied
here~\cite{Guh97} and will become more obvious when we discuss the
coset method in the next section. With the above notation we find
\begin{equation}
X_2(r,\lambda) \equiv X_2(\kappa,c) = 
\tilde{X}_2(\kappa,c) + \tilde{X}_2(-\kappa,c) + \kappa^2
\label{eq11}
\end{equation}
where
\begin{eqnarray}
\tilde{X}_2(\kappa,c) &=& \frac{2ic}{\pi} \int\limits_{-\infty}^\infty
\int\limits_{-\infty}^\infty
dx_1 dx_2 \frac{x_1x_2}{(x_1^2+x_2^2)^2} \exp(-{\cal L}(x_1,x_2)) 
\nonumber\\
{\cal L}(x_1,x_2) &=& \frac{1}{2c} \left[
\frac{x_1}{x_1+i} (x_1+i+\kappa)^2 + \frac{x_1+i}{x_1^+}
(x_2+i\kappa\frac{x_1}{x_1+i})^2 \right] \ .
\label{eq12}
\end{eqnarray}
To reduce the order of the pre-exponential singularity we perform an
integration by parts using $2x_2/(x_1^2+x_2^2)^2 = -\partial_{x_2}
(x_1^2+x_2^2)^{-1}$. With the additional transformation
$x_2 = y_2 x_1/(x_1+i)$ Eq.~(\ref{eq12}) takes the form
\begin{eqnarray}
\tilde{X}_2(\kappa,c) &=& -\frac{i}{\pi} 
\int\limits_{-\infty}^\infty
\int\limits_{-\infty}^\infty
dx_1 dy_2 \frac{(x_1+i)(y_2+i\kappa)}{(x_1+i)^2+y_2^2}
\exp(-\tilde{\cal L}(x_1,y_2))  \nonumber\\
\tilde{\cal L}(x_1,y_2) &=& \frac{1}{2c}\frac{x_1}{x_1+i}
 \left[ (x_1+i+\kappa)^2+(y_2+i\kappa)^2 \right] \ .
\label{eq13}
\end{eqnarray}
Now, our strategy is to eliminate the imaginary contributions to the
squares in $\tilde{\cal L}(x_1,y_2)$. To this end we perform two
shifts, namely $y_1 = x_1+i$ and $u_2 = y_2+i\kappa$. 
This amounts to moving the integration contour from the real axis to
the lines ${\rm Im}(x_1) = -1$ and ${\rm Im}(y_2) = -\kappa$,
respectively, see Fig.~\ref{fig1}. The integral (\ref{eq13}) remains
unchanged under these 
shifts unless the integrand exhibits singularities in the region
between the old and the new integration contours. 
With the help of Fig.~\ref{fig1} the reader can
easily convince himself that such singularities do indeed exist in our
case. They give rise to two different residuum contributions $R_1$ and
$R_2$. We therefore arrive at
\begin{eqnarray}
\tilde{X_2}(\kappa,c) &=& -\frac{i}{\pi} 
\int\limits_{-\infty}^\infty
\int\limits_{-\infty}^\infty
dy_1 du_2
\frac{y_1 u_2}{y_1^2 + (u_2-i\kappa)^2} \exp(-\tilde{\cal
  L}(y_1-i,u_2-i\kappa))  - R_1 - R_2   \nonumber\\
R_1 &=& \int\limits_{-1}^0 dy_2 \, (y_2+i\kappa) + \int\limits_0^1
dy_2 \, (y_2+i\kappa) 
\exp\left(2i\frac{\kappa}{c}(1-y_2)\right)  \nonumber\\
R_2 &=& \int\limits_0^{|\kappa|} dy_1 \, 
\left(\frac{\kappa}{|\kappa|}y_1-\kappa\right)
\exp\left(-(y_1-i)\frac{\kappa+|\kappa|}{c}\right) + 
\int\limits_{-|\kappa|}^0 dy_1 \, 
\left(\frac{\kappa}{|\kappa|}y_1+\kappa\right)
\exp\left(-(y_1-i)\frac{\kappa-|\kappa|}{c}\right)  \ .
\label{eq14}
\end{eqnarray}
We notice the appearance of ${\rm sign}(\kappa) = \kappa/|\kappa|$
which is needed to distinguish between the two cases $\kappa<0$ and
$\kappa>0$ in Eq.~(\ref{eq11}).  The residuum contributions can be
easily calculated analytically and with the final (innocuous)
transformation $u_1 = y_1+\kappa$ we obtain the following result for
the two--point level correlation function (\ref{eq11}),
\begin{eqnarray}
X_2(r,\lambda) &=& 1 +\frac{1}{2(\pi r)^2}
\left[ \exp\left(-2\frac{r^2}{\lambda^2}\right)\cos(2\pi r) -
  1 \right] 
+\frac{1}{(\pi\lambda)^2}  \nonumber\\
&& - \Bigg\{ \frac{i}{\pi} \int\limits_{-\infty}^\infty
\int\limits_{-\infty}^\infty
du_1 du_2 \frac{(u_1-\kappa)u_2}{(u_1-\kappa)^2+(u_2-i\kappa)^2}
\exp\left(-\frac{1}{2c}\left[ 1- \frac{i}{u_1-\kappa} \right]
(u_1^2+u_2^2) \right)
+ (\kappa \leftrightarrow -\kappa) \Bigg\}
\label{eq15}
\end{eqnarray}
In the limit $\lambda\to\infty$, the first two terms in
Eq.~(\ref{eq15}) reduce to the two--point level correlation function
for the GUE, 
\begin{equation}
X_2^{\rm GUE}(r) = 1 - \left(\frac{\sin \pi r}{\pi r}\right)^2 \ .
\label{eq15a}
\end{equation}
The additional terms are corrections for finite $\lambda$.
We denote the double integral in Eq.~(\ref{eq15}) by $P(c,\kappa)$. It
can be evaluated numerically for a wide range of values of $c$ and
$\kappa$ without too much effort. For our present purposes, in
particular for the comparison with the coset method in
Sec.~\ref{sec3}, we prefer to derive some analytical results  in the
limit $c\ll 1$.

\subsection{Perturbation theory for $X_2(r,\lambda)$}
\label{sec2_3}

In the following we assume that $c\to 0$, but $\kappa = {\rm const.}$
This means that the Breit--Wigner width is much larger than the level
spacing, $\Gamma_1 \gg 1$, and that the energy difference $r$ scales
like $\Gamma_1$.  Under these circumstances $P(c,\kappa)$ can be
calculated perturbatively, resulting in a power series in $c$.  With
the polar coordinates $u_1 = \rho \sin\varphi$ and $u_2 = \rho
\cos\varphi$ and after expanding both the exponent and the
pre-exponential term in Eq.~(\ref{eq15}) in powers of $\rho$ we obtain
the following expression for the double integral,
\begin{eqnarray}
P(c,\kappa) &=& \frac{1}{2\pi} \int\limits_0^{2\pi} d\varphi
\int\limits_0^\infty d\rho \, \rho e^{i\varphi} \cos\varphi
  \left(1-\frac{\rho}{\kappa}\sin\varphi\right) 
\sum_{m=0}^\infty \left(\frac{\rho e^{i\varphi}}{2\kappa i}\right)^m 
\exp\left(-\frac{i\rho^2}{2c\kappa}\sum_{n=1}^\infty
\left(\frac{\rho\sin\varphi}{\kappa}\right)^n \right) \times \nonumber\\
& &\times \exp\left(-\frac{1}{2}\frac{\kappa+i}{c\kappa} \rho^2
\right) \ . 
\label{eq17}
\label{EQ17}
\end{eqnarray}
The structure of the angular, i.e.~the $\varphi$, integration indicates
that only terms containing odd powers of $\rho$ lead to non-vanishing
contributions to $P(c,\kappa)$. From Appendix~\ref{app1} it is clear
that $P(c,\kappa)$ comprises
terms of order $c^1$, $c^2$, $c^3$, and so forth. For later reference,
we will calculate the $c^1$ and $c^2$ contribution explicitly.
To first order in $c$ the
contribution of the double integral $P(c,\kappa)$ to the two--point
correlator $X_2(r,\lambda)$ is (cf. Appendix~\ref{app1})
\begin{equation}
P^{(1)}(c,\kappa) + P^{(1)}(c,-\kappa)
= \frac{c\kappa}{2(\kappa+i)}
+ \frac{c\kappa}{2(\kappa-i)}
= \frac{1}{\pi}\frac{\Gamma_1}{r^2+\Gamma_1^2}
\left(\frac{r}{\Gamma_1}\right)^2  \ ,
\label{eq18}
\end{equation}
where we have used that $\Gamma_1 = \pi\lambda^2$. This result can be
combined with the term $1/(\pi^2\lambda^2) = 1/(\pi\Gamma_1) \equiv
c$ in Eq.~(\ref{eq15}) to give the full first--order contribution 
$X^{(1)}_2(r,\lambda)$ to $X_2(r,\lambda)$,
\begin{equation}
X^{(1)}_2(r,\lambda ) = \frac{1}{\pi}
\frac{\Gamma_1}{r^2+\Gamma_1^2} 
\label{eq19}
\end{equation}
The second--order contribution $X^{(2)}_2(r,\lambda)$ is composed
of four different combinations of terms from the
perturbation series~(\ref{eq17}), see Appendix~\ref{app1}. Combined
they can be written as
\begin{eqnarray}
X^{(2)}_2(r,\lambda) &=& \frac{c^2}{4} \left[ \frac{1}{(\kappa+i)^2} +
\frac{2i}{(\kappa+i)^3} + \frac{3}{(\kappa+i)^4} 
 \quad + (\kappa \leftrightarrow -\kappa) \quad \right] \nonumber\\
&=& -\frac{1}{2\pi^2} \frac{\Gamma_1^2 - r^2}{(r^2+\Gamma_1^2)^2}
+ \frac{c^2}{4} \left[\frac{2i}{(\kappa+i)^3} + \frac{3}{(\kappa+i)^4} 
 \quad + (\kappa \leftrightarrow -\kappa) \quad \right] \ .
\label{eq20}
\end{eqnarray}
In the last line of the above equation we have distinguished between
two contributions to $X^{(2)}_2(r,\lambda)$, a first piece
originating from the term with the quadratic denominator, and ``extra''
contributions arising from the terms with higher--order denominators.
The significance of this distinction will become clear in the
following section, where we essentially re-derive the present results
from the point of view of the coset method.

\section{Coset method}
\label{sec3}

The supersymmetric ``coset'' method~\cite{Efe83,Ver85} has been widely
employed in the last ten years or so to solve problems of random
matrix theory~\cite{Guh97a}. In the present context, it was already
used in Ref.~\cite{Guh89} for a qualitative discussion. To make
contact with the well--established methods used in the literature over
the years we will adopt normalization conventions in this section
which differ slightly from those used in Sec.~\ref{sec2}.

\subsection{Basic definitions}
\label{sec3_1}

We write the full Hamiltonian $H$ in
Eq.~(\ref{eq1}) as 
\begin{equation}
H = H_0 + H_1 \ ,
\label{eq21}
\end{equation}
thereby effectively absorbing the strength parameter $\alpha$ into the
definition of $H_1$. The distribution $p^{(0)}(\eta)$ of the
non-vanishing elements $H_0^{ii}$ of $H_0$ defines an important energy
scale, the bandwidth $B$, because we have $p^{(0)}(H_0^{ii}) \propto
B^{-1}$ for reasons of normalization.  The probability distribution
function $P^{(1)}_N(H_1)$ adopted here is a slightly modified version of
Eq.~(\ref{eq4}) and reads
\begin{equation}
P^{(1)}_N(H_1) = 2^{N(N-1)/2}
\left(\frac{N}{2\pi\gamma^2}\right)^{N^2/2}
  \exp\left(-\frac{N}{2\gamma^2}{\rm tr} H_1^2 \right)
\quad \Longrightarrow \quad
\langle \vert H_1^{ij}\vert^2\rangle = \frac{\gamma^2}{N} \ ,
\label{eq22}
\end{equation}
where the symbol $\langle \ldots\rangle$ denotes averaging over the
$H_1$--ensemble (i.e. the GUE). The parameter $\gamma$ introduced in
this way is related to the strength parameter $\lambda$ employed in
Sec.~\ref{sec2}  through
\begin{equation}
\lambda = \sqrt{\frac{2}{N}}\frac{\gamma}{D} \ .
\label{eq22a}
\end{equation}

It is our purpose in this section to calculate the two--point
correlator $X_2$ (and later the number variance $\Sigma^2$) for the
Hamiltonian~(\ref{eq21}) perturbatively in a certain suitable range of
parameters. To define this range we note that $D=B/N$ is the level
spacing of $H_0$ and $\gamma/\sqrt{N}$ the typical strength of the
perturbing matrix elements. Hence we find for the (dimensionless)
induced spreading or Breit--Wigner width $\Gamma_1$
\begin{equation}
\Gamma_1 D\propto \frac{\gamma^2}{ND} = \frac{\gamma^2}{B} \ .
\label{eq23}
\end{equation}
Our calculation in this section is valid under the two conditions
\begin{equation}
\frac{\gamma}{B} = \sqrt{\frac{\Gamma_1 D}{B}} \to 0
 \ , \qquad
\Gamma_1 \propto \frac{\gamma^2}{B D} \gg 1 \qquad\qquad \mbox{(but
  finite)} \ .
\label{eq24}
\end{equation}
The first of the conditions~(\ref{eq24}) means that the bandwidth $B$
is infinitely larger than the Breit--Wigner width $\Gamma_1 D$. This
ensures that neither $B$ nor the level spacing $D$ are appreciably
changed by the perturbation $H_1$. The second condition ensures that
we deal with overlapping resonances, i.e. the original states of $H_0$
are thoroughly mixed to form the new eigenstates of the combined
Hamiltonian $H$. Our perturbative calculation relies on the small
parameter $1/\Gamma_1$. Again, as in the previous section, the
parameter $\kappa = r/\Gamma_1$ is held fixed so that the
dimensionless energy difference $r$ scales like $\Gamma_1$.

With the Green function $G_{\pm} = (x^\pm - H)^{-1}$ the density of
states is given by
\begin{equation}
\rho(x) = \frac{1}{2\pi} ({\rm tr} \, G_+ - {\rm tr} \, G_- ) 
\label{eq25}
\end{equation}
and hence the averaged density--density correlation function takes the
form 
\begin{eqnarray}
\rho_2\left(x+\omega/2,x-\omega/2\right) 
&=& \
\big[\big\langle \rho\left(x+\omega/2\right)
\rho\left(x-\omega/2\right) \big\rangle \big]_0 \nonumber\\
&=& \frac{1}{2\pi^2} {\rm Re} \big[\big\langle {\rm tr}\, 
G_+\left(x+\omega/2\right)
\left( {\rm tr}\, G_-\left(x-\omega/2\right) - 
{\rm tr}\, G_+\left(x-\omega/2\right) \right)
\big\rangle\big]_0 \ .
\label{eq26}
\end{eqnarray}
Here, we have introduced the notation $[\ldots]_0$ for the
$H_0$--average. The angular brackets $\langle\ldots\rangle$ denote, as
above, averaging over $H_1$.  
In the sequel we will typically perform the
$H_1$--average in a rather early step of the calculation while the
$H_0$--average is performed in the final stages. We note that
\begin{equation}
\rho_2(x+\omega/2,x-\omega/2) = \frac{1}{D^2} X_2(r) +
\frac{1}{D^2}\delta(r) 
\label{eq27}
\end{equation}
defines the relation of the density--density correlator to the
spectral two--point function (\ref{eq9}) considered earlier. Here
$r=\omega/D$ is the dimensionless energy difference as in
Sec.~\ref{sec2}. 

\subsection{One--point function and density of states}
\label{sec3_2}

To illustrate our procedure and to define relevant energy scales we
start with the average one--point function. With the supersymmetric
generating functional
\begin{eqnarray}
Z_\pm(q_\pm) &=& {\rm detg}^{-1}([x^{\pm} - H]\otimes 1_2
+1_N\otimes q_\pm L_g/2) \nonumber\\
&=& \exp\left(-{\rm trg}\ln\left(1_N\otimes 1_2 +
G_\pm \otimes q_\pm L_g/2 \right)\right)
\label{eq28}
\end{eqnarray}
we can write the Green function as
\begin{equation}
{\rm tr}\, G_\pm(x) = \frac{\partial}{\partial_{q_\pm}}
Z_\pm(q_\pm)\bigg\vert_{q=0} \ , 
\label{eq29}
\end{equation}
where $L_g = {\rm diag}(-1,1)$ is a two--dimensional (super--) matrix.
With standard techniques~\cite{Efe83,Ver85} the $H_1$--average of the
generating functional can be easily calculated to give
\begin{equation}
\big\langle Z_+(J) \big\rangle = \int d[\sigma]
\exp\left(-\frac{N}{2} {\rm trg}(\sigma^2) - {\rm trg}\ln(
     [x^+-H_0]\otimes 1_2 + 1_N\otimes[\gamma\sigma + J])
   \right) \ .
\label{eq30}
\end{equation}
We have introduced the source matrix $J = qL_g/2$, and $\sigma$ is a
$2\times 2$ supermatrix as in Eq.~(\ref{eq6}). In contradistinction to
the previous section the further evaluation of Eq.~(\ref{eq30}) relies
on the saddle--point approximation. With the saddle--point equation
\begin{equation}
\sigma = -\frac{\gamma}{N} \sum_j \frac{1}{x^+-H_0^{jj} +
  \gamma\sigma}
\label{eq31}
\end{equation}
and with the ansatz $\gamma\sigma = (\Gamma_0+i\Gamma_1)D/2 = \Gamma
D/2$ we obtain
\begin{eqnarray}
\Gamma &=& -\frac{2\gamma^2}{ND} \sum_j \frac{1}{x^+-H_0^{jj} +
  D\Gamma/2} \nonumber\\
&\approx& -\frac{2\gamma^2}{D} \int d\eta\ 
  \frac{p^{(0)}(\eta)}{x^+-\eta+D\Gamma/2} 
  \ .
\label{eq32}
\end{eqnarray}
Equation~(\ref{eq32}) can be solved in the limit $|\Gamma|\ll B$ and
we arrive at the approximate expressions
\begin{eqnarray}
\Gamma_0 &\approx& -\frac{2\gamma^2}{D} {\bf P} \int d\eta\ 
\frac{p^{(0)}(\eta)}{x-\eta} \nonumber\\
\Gamma_1 &\approx& 2\pi\frac{\gamma^2}{D} p^{(0)}(0) =
2\pi\frac{\gamma^2}{BD} 
\qquad\qquad \mbox{(Breit--Wigner width)} \ .
\label{eq33}
\end{eqnarray}
As immediate consequences we find
\begin{eqnarray}
\big\langle {\rm tr} \, G_+\big\rangle &=& 
{\rm tr} \frac{1}{x^+-H_0+D\Gamma/2} = 
-\frac{N\Gamma D}{2\gamma^2}
 \nonumber\\
\big\langle \rho(x) \big\rangle &=& -\frac{1}{\pi} {\rm Im}
\big\langle {\rm tr} \, G_+ \big\rangle_1 = \frac{N\Gamma_1 D}{2\pi\gamma^2}
 = N p^{(0)}(x) \ .
\label{eq34}
\end{eqnarray}
In particular, the last equation means that the $H_0$ average of the
mean level density can be trivially performed, giving $[\langle
\rho(x) \rangle]_0 = N p^{(0)}(x)$.

We see from Eq.~(\ref{eq26}) that we need the average of a $G_+G_-$
and of a $G_+G_+$ term.
It is well known that in the large $N$ limit the $H_1$--average of the
product of two Green functions with infinitesimal increments of equal
sign factorizes,
\begin{equation}
\big\langle {\rm tr}\, G_+(x+\omega/2) {\rm tr}\, G_+(x-\omega/2)
\big\rangle = 
\big\langle {\rm tr}\,G_+ \big\rangle \big\langle{\rm tr}\,G_+
\big\rangle =
{\rm tr} \frac{1}{x^++\omega/2-H_0+D\Gamma/2}
{\rm tr} \frac{1}{x^+-\omega/2-H_0+D\Gamma/2} \ .
\label{eq35}
\end{equation}
The same turns out to be true after averaging over $H_0$. 
This average is easily performed by simply integrating over the
independent random entries of the diagonal matrix $H_0$,
\begin{eqnarray}
\big[\big\langle {\rm tr}\, G_+ {\rm tr}\, G_+
\big\rangle\big]_0 &\approx& N(N-1) 
\int d\eta\
\frac{p^{(0)}(\eta)}{x^++\omega/2 - \eta + D\Gamma/2}
\int d\eta\
\frac{p^{(0)}(\eta)}{x^+-\omega/2 - \eta + D\Gamma/2}
\nonumber\\ 
&& + N\int d\eta\ \frac{p^{(0)}(\eta)}{
\left(x^++\omega/2 - \eta + D\Gamma/2\right)
\left(x^+-\omega/2 - \eta + D\Gamma/2\right)} \nonumber\\
&=& \frac{N(N-1)}{N^2} 
\big[\big\langle {\rm tr}\, G_+
\big\rangle\big]_0
\big[\big\langle {\rm tr}\, G_+
\big\rangle\big]_0 
\approx
\big[\big\langle {\rm tr}\, G_+
\big\rangle\big]_0
\big[\big\langle {\rm tr}\, G_+
\big\rangle\big]_0  \ .
\label{eq36}
\end{eqnarray}
The term $\propto N$ in Eq.~(\ref{eq36}) is seen to vanish upon
closing the contour in the lower half plane. We conclude that in order 
to obtain the density--density correlator (\ref{eq26}) we can simply 
replace $\langle \mbox{tr}\,G_\pm\rangle=\mp i\pi/D$. 

\subsection{Two--point function and nonlinear $\sigma$ model}
\label{sec3_3}

Our next goal is to calculate the remaining quantity $\langle {\rm tr}
\, G_+ {\rm tr}\, G_- \rangle$, which leads to the familiar nonlinear
$\sigma$ model, and its $H_0$--average. First, we generalize the
definition of the supersymmetric generating functional (\ref{eq30}),
\begin{eqnarray}
\big\langle {\rm tr}\, G_+(x+\omega/2) {\rm tr}\,
G_-(x-\omega/2)\big\rangle &=& 
\frac{\partial}{\partial_{q_+}}
\frac{\partial}{\partial_{q_-}}
\big\langle Z(q_+,q_-)\big\rangle \bigg\vert_{q_\pm =0} =
\frac{\partial}{\partial_{q_+}}
\frac{\partial}{\partial_{q_-}}
\int d[Q] \exp\left( - {\cal L}(Q)\right)
\bigg\vert_{q_\pm =0}   \nonumber\\
{\cal L}(Q) &=& \sum_j {\rm trg}\ln \left( [x-H_0^{jj}]1_4 +
\frac{\omega}{2} \Lambda + \frac{D}{2}\Gamma Q + J \right) \ .
\label{eq37}
\end{eqnarray}
The quantity $Q$ is a $4\times 4$ supermatrix parameterizing the
saddle--point manifold~\cite{Efe83,Ver85} to which the integration has
been restricted by the saddle--point approximation. Furthermore,
$J={\rm diag}(q_+L_g,q_-L_g)/2 = q_+ P_+ + q_- P_-$ (this defines the
projectors $P_+$ and $P_-$), and $\Lambda = {\rm diag}(1_2,-1_2)$. To
perform the derivatives in Eq.~(\ref{eq37}) it is useful to expand
${\cal L}$ up to second order in the source matrix $J$,
\begin{equation}
{\cal L}(Q) = {\cal L}_0(Q) + {\cal L}_1(Q) + {\cal L}_2(Q) + \ldots
\label{eq38}
\end{equation}
The three contributions can be written as
\begin{eqnarray}
{\cal L}_0 &=& \sum_j {\rm trg}\ln [ (x-H_0^{jj})1_4 +
(\omega\Lambda + \Gamma_0 D + i\Gamma_1 D Q)/2]   \nonumber\\
{\cal L}_1 &=& \sum_j {\rm trg} [ g_j J] \nonumber \\
{\cal L}_2 &=& -\frac{1}{2}\sum_j {\rm trg}[g_jJg_jJ] \ ,
\label{eq39}
\end{eqnarray}
where we have introduced the abbreviation $g_j =
[x+(\omega/2)\Lambda-H_0^{jj} + (\Gamma_0 D+i\Gamma_1 D
Q)/2]^{-1}$. After performing the source term derivatives in
Eq.~(\ref{eq37}) we arrive at a central equation of this section,
\begin{eqnarray}
\big\langle {\rm tr}\, G_+(x+\omega/2) 
{\rm tr}\, G_-(x-\omega/2)
\big\rangle &=& \int d[Q] (S_1(Q) + S_2(Q)) \exp(-{\cal L}_0(Q))
\nonumber \\
S_1(Q) &=& \sum_{j,k} {\rm trg}[g_jP_+] {\rm trg}[g_k P_-] \nonumber\\
S_2(Q) &=& \sum_j {\rm trg} [g_jP_+ g_jP_-] \ .
\label{eq40}
\end{eqnarray}
This is the particular form of the zero--dimensional nonlinear
$\sigma$ model describing the crossover between Poisson and GUE
statistics.

\subsection{Evaluation of the nonlinear $\sigma$ model}
\label{sec3_4}

As already mentioned above, the $Q$--integration in Eq.~(\ref{eq40})
is restricted to the saddle--point manifold familiar from numerous
previous applications of the nonlinear $\sigma$
model~\cite{Guh97a}. It belongs to the peculiar features of
superanalysis that integrals of the type (\ref{eq40}), which are
derived by a supersymmetric change of variables, contain an extra
``boundary'' contribution (sometimes referred to as the Efetov--Wegner
term). This boundary contribution is generically given by the value of
the integrand at $Q=\Lambda$. In our present case we have
${\cal L}_0(\Lambda) = 0$, $S_2(\Lambda)=0$, and
\begin{equation}
S_1(\Lambda) = \sum_{j,k}
\frac{1}{x-H_0^{jj}+D\Gamma_0/2 + (\omega+i\Gamma_1 D)/2} \
\frac{1}{x-H_0^{kk}+D\Gamma_0/2 - (\omega+i\Gamma_1 D)/2} \ .
\label{eq41}
\end{equation}
The $H_0$--average is performed as in Eq.~(\ref{eq36}),
\begin{eqnarray}
S_1(\Lambda) &=& N(N-1) 
\int d\eta \ \frac{p^{(0)}(\eta)}{
x-\eta+D\Gamma_0/2 + (\omega+i\Gamma_1 D)/2}
\int d\eta \ \frac{p^{(0)}(\eta)}{
x-\eta+D\Gamma_0/2 - (\omega+i\Gamma_1 D)/2} +
\nonumber\\ 
&& + N\int d\eta\ \frac{p^{(0)}(\eta)}{
  \left(x-\eta+D\Gamma_0/2 +
  (\omega+i\Gamma_1 D)/2\right) \left(x-\eta+D\Gamma_0/2 -
  (\omega+i\Gamma_1 D)/2\right)}  \nonumber\\
&\approx& \big[\big\langle {\rm tr}\,
G_+(x+\omega/2)\big\rangle\big]_0 
\big[\big\langle
{\rm tr}\, G_-(x-\omega/2) \big\rangle\big]_0
+ \frac{1}{D} \int dz \frac{1}{z^2 - (\omega+i\Gamma_1 D)^2/4}
\nonumber\\ 
&=&
\big[\big\langle {\rm tr}\,
G_+(x+\omega/2)\big\rangle\big]_0 
\big[\big\langle
{\rm tr}\, G_-(x-\omega/2) \big\rangle\big]_0
+\frac{2\pi}{D^2} \frac{1}{\Gamma_1-ir} \ .
\label{eq42}
\end{eqnarray}
We recall that $r = \omega/D$.
The first term in Eq.~(\ref{eq42} together with Eq.~(\ref{eq36}) forms
the disconnected contribution to the two--point correlation function
$X_2(r)$. Upon inserting the results (\ref{eq33}) and (\ref{eq34}) one
can easily show that the sum of disconnected contributions to $X_2(r)$
reduces to unity (for $\Gamma_0\approx 0$). 
The second term in Eq.~(\ref{eq42}) is part of the connected
contributions. In the following we calculate the remaining connected
terms by treating the saddle--point integration perturbatively.

We start from Eq.~(\ref{eq40}). The general strategy will be to
express both the exponent and the pre-exponential terms in
Eq.~(\ref{eq40}) in terms of the independent variables of the
saddle--point manifold. It turns out that we have to consider only
quadratic terms in the exponent so that the integration becomes
Gaussian and therefore trivial. Concerning the $H_0$--average it is
important to note that 
those terms involving correlations between the matrix elements of
$H_0$ in the exponent and in the pre-exponential terms can be
neglected. In fact, it turns out that the pre-exponential terms and the
exponent can be {\it independently} averaged over $H_0$. This amounts
to a tremendous simplification of our calculation.
A more thorough discussion of these issues as well as a
number of technical steps omitted here for clarity can be found in
Appendix~\ref{app2}.

We express the deviation of $Q$ from the diagonal value $\Lambda$ in
terms of the quantity $\Delta Q = \{\Lambda,Q\}/2-1$, where $\{.,.\}$
denotes the anticommutator. Then the $H_0$--average of ${\cal L}_0(Q)$
can be expressed as (see Appendix~(\ref{app2}))
\begin{equation}
\big[ {\cal L}_0(Q) \big]_0 =
-i\frac{\pi}{2}\frac{r\Gamma_1}{\Gamma_1-ir}
{\rm trg}(\Delta Q) \ .
\label{eq44}
\end{equation}
For our present purposes, a suitable parameterization of the
saddle--point manifold is given by
\begin{eqnarray}
Q &=& T^{-1}\Lambda T = \Lambda T^2 \nonumber\\
T &=& \sqrt{1+R^2} + R \nonumber\\
R &=& \left[ \begin{array}{cc}
        0 & t \\
        \overline{t} & 0 
      \end{array} \right] \ ,
\label{eq45}
\end{eqnarray}
where $t$ and $\overline{t}$ are $2\times 2$ supermatrices
representing the unrestricted ``free'' variables of the saddle--point
manifold. It follows for ${\cal L}_0$ that
\begin{equation}
\big[ {\cal L}_0(Q) \big]_0 =
-2\pi i \frac{r\Gamma_1}{\Gamma_1-ir} 
{\rm  trg}(\overline{t}t) \ .
\label{eq46}
\end{equation}
Likewise we have to express the $H_0$--averages of $S_1$ and $S_2$ in
terms of $t$ and $\overline{t}$. Again, the detailed derivation has
been deferred to Appendix~\ref{app2},
\begin{eqnarray}
\big[ S_1 \big]_0 &=& 4
\frac{\pi^2}{D^2} \frac{\Gamma^4_1}{(\Gamma_1-ir)^4}  {\rm
  trg}[t\overline{t}P_+]{\rm trg}[\overline{t}tP_-] \nonumber\\
\big[ S_2 \big]_0 &=& -4\frac{\pi}{D^2}
\frac{(i\Gamma_1)^2}{(\Gamma_1-ir)^3}  {\rm trg}[tP_+\overline{t}
P_-] \ .
\label{eq47}
\end{eqnarray}
Using Eqs.~(\ref{eq40}), (\ref{eq46}), and (\ref{eq47})
the remaining Gaussian integrations can be easily performed. To this
end it is useful to employ certain Wick--type contraction rules as
explained in Ref.~\onlinecite{Iid90}. Along these lines we obtain the
following ``diffusion'' 
contribution from a perturbative treatment of the saddle--point
integral,
\begin{equation}
\big[\big\langle {\rm tr}\,G_+ {\rm tr}\,G_-
\big\rangle\big]_0^{\rm diff} 
=
-\frac{1}{(Dr)^2}\frac{\Gamma^2_1}{(\Gamma_1-ir)^2} +
\frac{2i\Gamma_1}{D^2r(\Gamma_1-ir)^2} 
= -\frac{1}{(Dr)^2} - \frac{1}{D^2(\Gamma_1-ir)^2} \ .
\label{eq48}
\end{equation}
The individual contributions of $S_1$ and $S_2$ to this result
correspond to the first and the second term on the r.h.s. of the first
line, respectively.

In summary, the total ($H_0$ and $H_1$) average of ${\rm tr}G_+{\rm
  tr}G_-$ consists of three terms, the disconnected part, the boundary
  term (\ref{eq42}), and the diffusion contribution (\ref{eq48}),
\begin{equation}
\big[\big\langle {\rm tr}\,G_+ {\rm tr}\,G_-
\big\rangle\big]_0 =  C_{\rm disc} + C_{\rm bound} + C_{\rm
  diff} \ .
\label{eq49}
\end{equation}
By definition, see Eqs.~(\ref{eq26}) and (\ref{eq27}), we have
\begin{equation}
X_2(r) = \frac{D^2}{2\pi^2} {\rm Re} ( C_{\rm bound} + C_{\rm
  diff} ) \ .
\label{eq50}
\end{equation}
If we insert our results for $C_{\rm bound}$ and $C_{\rm diff}$ and
generalize the latter to arbitrary (GUE, GOE, GSE) symmetry by
multiplying it with $2/\beta$ ($\beta = 1,2,4$) we arrive at
\begin{eqnarray}
X_2(r) &=& 1 -\frac{1}{\beta}  \frac{1}{(\pi r)^2} +
X_2^{(1)}(r) + X_2^{(2)}(r)      \nonumber\\
X_2^{(1)}(r) &=&
\frac{1}{\pi} \frac{\Gamma_1}{\Gamma^2_1+r^2} \nonumber\\
X_2^{(2)}(r) &=&
-\frac{1}{\pi^2\beta}
\frac{\Gamma^2_1-r^2}{(\Gamma^2_1+r^2)^2}  \ .
\label{eq51}
\end{eqnarray}
The first term in this result is the perturbative (divergent)
expression for the two--point correlation function of the Gaussian
ensembles. It can be replaced by the full non-perturbative result, i.e.
by Eq.~(\ref{eq15a}) in the case of the GUE. Comparison of
the first and second order contributions to $X_2(r)$,
$X_2^{(1)}(r)$ and  $X_2^{(2)}(r)$, obtained so far with the
corresponding results of the graded eigenvalue method in Eqs.~(\ref{eq19})
and (\ref{eq20}), reveals missing terms in the present calculation. We
conclude this section with a discussion of the origin of this
discrepancy.

\subsection{Fluctuating Breit--Wigner width}
\label{sec3_5}

Let us recall the saddle--point equation (\ref{eq32}). From the first
line of this equation it is clear that $\Gamma$ depends on the
particular realization of $H_0$. But so far we have only considered
one fixed (mean) value of $\Gamma$ and have neglected random
fluctuations of $\Gamma$ within the ensemble. We will see below that
these random 
fluctuations of the Breit--Wigner width lead to the missing terms in
our expression for $X_2(\omega)$. To discuss the effect of
fluctuations we choose the distribution of the (diagonal) matrix
elements of $H_0$ to be
\begin{equation}
p^{(0)}(\eta) = \frac{1}{2W} \Theta(W-|\eta|)
\label{eq52}
\end{equation}
for definiteness. This means that the bandwidth $B$ is given by
$B=2W$. Then, according to Eq.~(\ref{eq33}),
\begin{eqnarray}
\Gamma_0 &=& 0           \nonumber\\
\Gamma_1 &=& \frac{2\pi\gamma^2}{ND^2}  \quad\quad (D=2W/N) \ .
\label{eq53}
\end{eqnarray}
We note that if we insert the relation (\ref{eq22a}) between $\gamma$ and
$\lambda$ in Eq.~(\ref{eq53}) we obtain $\Gamma_1 = \pi\lambda^2$ as
claimed below Eq.~(\ref{eq10a}).
In the approximation (\ref{eq53}) $\Gamma$ is given by
\begin{equation}
\Gamma^{(0)} = i\Gamma_1 \ .
\label{eq54}
\end{equation}
With this notation the exact saddle--point equation (\ref{eq32}) reads
\begin{equation}
\Gamma = -\frac{\Gamma_1 D}{\pi} \sum_j \frac{1}{x-H_0^{jj}
  +D\Gamma/2}  \ .
\label{eq55}
\end{equation}
If we insert our lowest (zeroth) order approximation (\ref{eq54}) on
the r.h.s. of Eq.~(\ref{eq55}) we obtain the next (first) order in an
iterative refinement procedure,
\begin{equation}
\Gamma^{(1)} = -\frac{\Gamma_1 D}{\pi} \sum_j \frac{1}{x-H_0^{jj}
  +D\Gamma^{(0)}/2} \ ,
\label{eq56}
\end{equation}
so that the random fluctuations of $\Gamma_c$ are given by
\begin{equation}
\delta\Gamma =  \Gamma^{(1)} - \Gamma^{(0)} = 
-\frac{\Gamma_1 D}{\pi}\sum_j \frac{1}{x-H_0^{jj}  +iD\Gamma_1/2}  -
i\Gamma_1 \ . 
\label{eq57}
\end{equation}
These fluctuations lead to additional terms in the boundary
contribution because their inclusion modifies the r.h.s. of
Eq.~(\ref{eq41}). With the definitions
\begin{eqnarray}
G_j &=& \frac{1}{x+\omega/2 - H_0^{jj} +
  D\Gamma^{(0)}/2} \nonumber\\ 
\overline{G}_j &=& \frac{1}{x-\omega/2 - H_0^{jj} +
  D\Gamma^{(0)*}/2}
\label{eq58}
\end{eqnarray}
we have
\begin{eqnarray}
\frac{1}{x+\omega/2 - H_0^{jj} +
  D(\Gamma^{(0)}+\delta\Gamma)/2} 
&\approx& G_j - \frac{D}{2} \delta\Gamma G_j^2 \nonumber\\
\frac{1}{x-\omega/2 - H_0^{jj} +
  D(\Gamma^{(0)*}+\delta\Gamma^*)/2} 
&\approx& 
\overline{G}_j - \frac{D}{2} \delta\Gamma^* \overline{G}_j^2 \ .
\label{eq59}
\end{eqnarray}
Hence we obtain instead of Eq.~(\ref{eq41})
\begin{eqnarray}
S_1(\Lambda) &=& \sum_j \left(G_j - \frac{D}{2}\delta\Gamma  G_j^2\right)
                 \sum_k \left(\overline{G}_k - \frac{D}{2}\delta\Gamma^* 
                         \overline{G}_k^2\right)   \nonumber\\
\delta\Gamma &=& -\frac{\Gamma_1 D}{\pi} \sum_l G_l -i\Gamma_1
                 \nonumber\\
\delta\Gamma^* &=& -\frac{\Gamma_1 D}{\pi} \sum_l \overline{G}_l
                 +i\Gamma_1  \ .
\label{eq60}
\end{eqnarray}
Averaging $S_1(\Lambda)$ over $H_0$ is tedious but in principle
straightforward (see Appendix~\ref{app3}). The result of this
calculation is that there are additional contributions to
Eq.~(\ref{eq42}),
\begin{equation}
C_{\rm bound}^{\rm extra} = -\frac{2\Gamma_1}{D^2(\Gamma_1-ir)^3} + 
\frac{3\Gamma_1^2}{D^2(\Gamma_1-ir)^4},
\label{eq61}
\end{equation}
leading to the extra terms
\begin{equation}
X_{2, \rm extra}^{(2)}(\omega) = \frac{c^2}{4} \left[
\frac{2i}{(\kappa+i)^3} + \frac{3}{(\kappa+i)^4} + (\kappa
\leftrightarrow -\kappa) \right]
\label{eq62}
\end{equation}
in the second order contribution to $X_2(\omega)$. These are precisely
the missing terms in Eq.~(\ref{eq51}). We conclude that random
fluctuations of the Breit--Wigner width manifest themselves in the
two--point level correlation function. These extra contributions are
automatically included in the exact graded eigenvalue method.

\section{Number Variance and Numerical Simulations}
\label{sec4}

The number variance $\Sigma^2(r)$ for a random matrix ensemble is
defined as the variance of the number of energy levels within a
dimensionless energy interval of fixed length $r$. It can be easily
calculated from the spectral two--point correlation function
$X_2(r,\lambda)$,
\begin{eqnarray}
\Sigma^2(r) &=& r - 2\int\limits_0^r (r-r')(1-X_2(r',\lambda)) dr' 
                             \nonumber\\
       &=& -r(r-1) + 2\int\limits_0^r (r-r')X_2(r',\lambda) dr' \ .
\label{eq4:1}
\end{eqnarray}
This spectral observable is particularly suitable to identify the
energy scale at which the spectral statistics of the ensemble
(\ref{eq1}) deviates from either the GUE or Poisson behavior. For the
Poisson ensemble, $\Sigma^2(r)$ is simply given by the straight
line, $\Sigma^2_{\rm Poi}(r) = r$, while for the GUE the number
variance is of 
roughly logarithmic shape. For intermediate cases, i.e. for finite
non-vanishing $\alpha$ in Eq.~(\ref{eq1}), one expects $\Sigma^2(r)$ to
exhibit GUE behavior for small energy separations $r$. For larger
energy intervals it should the cross over to a linear Poisson--like
form. In Sec.~\ref{sec4_1} we determine, based on Eq.~(\ref{eq4:1})
and the correlator $X_2(r,\lambda)$ calculated in the previous
sections, the energy scales relevant for this
crossover. Section~\ref{sec4_2} is then devoted to the question
whether our theoretical expectations are borne out by direct numerical
simulations.

\subsection{Crossover Energy Scales}
\label{sec4_1}

The principal result of the analysis performed in
Secs.~\ref{sec2} and \ref{sec3} for the case $\Gamma_1\gg 1$ 
is that the two--point level correlator
$X_2(r,\lambda)$ for the random matrix model (\ref{eq1}) 
can be written as 
\begin{eqnarray}
\label{eq4:2}
X_2(r,\lambda) &=& X_2^{\rm GUE}(r) + X_2^{\rm NA}(r,\lambda)+
X_2^{(1)}(r,\lambda) +X_2^{(2)}(r,\lambda) \\
\nonumber\\
\nonumber
X_2^{\rm GUE}(r) &=& 1 - \frac{\sin^2(\pi r)}{(\pi r)^2} 
=1-\frac{1}{2\,\pi^2\,r^2}\left(1-\cos(2\pi r)\right)\\
\nonumber\\
\nonumber
X_2^{\rm NA}(r,\lambda) &=& \frac{1}{2\,\pi^2\,r^2}
\left[\exp\left(-2\pi\frac{r^2}{\Gamma_1}\right)-1\right]\,\cos(2\pi r)\\
\nonumber\\
\nonumber
X_2^{(1)}(r,\lambda) &=& \frac{1}{\pi}
\frac{\Gamma_1}{\Gamma_1^2 + r^2} \\
\nonumber\\
\nonumber
X_2^{(2)}(r,\lambda) &=& 
-\frac{1}{2\pi^2}\frac{\Gamma_1^2-r^2}{(\Gamma_1^2 + r^2)^2}
+ \frac{1}{2\pi^2}\left(
\frac{-2\Gamma_1^2(\Gamma_1^2-3 r^2)}{(\Gamma_1^2+r^2)^3}+
\frac{3\Gamma_1^2(\Gamma_1^4-6\Gamma_1^2 r^2+r^4)}
{(\Gamma_1^2+r^2)^4}\right)
\end{eqnarray}
with $\Gamma_1=\pi\lambda^2=\pi(\alpha/D)^2$. The first contribution 
$X_2^{\rm GUE}(r)$ is just the two-point level correlator for the GUE 
case. The contribution $X_2^{\rm NA}(r,\lambda)$ describes nonanalytic 
corrections for $\Gamma_1^{-1}\ll 1$ (with $r$ of order $\Gamma_1$) 
that cannot be obtained by perturbation theory. They are found 
in the framework of the graded eigenvalue 
method or a more sophisticated nonperturbative evaluation of the 
coset method. This contribution tends to eliminate the oscillating 
part of $X_2^{\rm GUE}(r)$ for $r\gg \sqrt{\Gamma_1}$. 
The terms $X_2^{(1)}(r,\lambda)$ and $X_2^{(2)}(r,\lambda)$ are 
just the first or second order corrections in $\Gamma_1^{-1}$.
In (\ref{eq4:2}), we have focused on the unitary case 
($\beta = 2$) for definiteness. Our qualitative
conclusions, however, apply likewise to the orthogonal and the
symplectic symmetry class ($\beta=1,\,4$). The nonanalytic correction 
can be neglected when performing the $r'$-integral in (\ref{eq4:1}) to 
determine the number variance. As a consequence of the form of
Eq.~(\ref{eq4:2}) we obtain 
\begin{equation}
\Sigma^2(r,\lambda) = \Sigma^2_{\rm GUE}(r) +
\Sigma^2_{(1)}(r,\lambda) + \Sigma^2_{(2)}(r,\lambda) \ .
\label{eq4:3}
\end{equation}
A short calculation using Eqs.~(\ref{eq4:1}) and (\ref{eq4:2}) reveals
that
\begin{equation}
\Delta\Sigma^2(r,\lambda) = \Sigma^2_{(1)}(r,\lambda) +
\Sigma^2_{(2)}(r,\lambda) = \frac{2r}{\pi} {\rm
  arctan}\left(\frac{r}{\Gamma_1}\right)  - 
\left(\frac{\Gamma_1}{\pi}+\frac{1}{2\pi^2}\right)
\ln\left(1+\left(\frac{r}{\Gamma_1}\right)^2\right)   +
 \frac{1}{2\pi^2}\frac{r^2\Gamma_1^2 -
  r^4}{(r^2+\Gamma_1^2)^2} \ .
\label{eq4:4}
\end{equation}
It is consistent with  our assumption $\Gamma_1\gg 1$ to neglect the
second term $1/2\pi^2$ in the prefactor of the logarithmic
contribution and the last term in Eq.~(\ref{eq4:4}). To identify the
energy scales of 
interest we first consider the limit $r\ll \Gamma_1$. Then
\begin{equation}
\Delta\Sigma^2(r,\lambda) \approx \frac{2r^2}{\pi\Gamma_1} -
\frac{\Gamma_1}{\pi}
\left(\frac{r}{\Gamma_1}\right)^2 =
\frac{r^2}{\pi\Gamma_1} \qquad\qquad (r\ll\Gamma_1) \ .
\label{eq4:5}
\end{equation}
Obviously, the scale set for deviations from the GUE behavior is
$\sqrt{\Gamma_1}$ in agreement with previous findings~\cite{Guh97}. 
    
Let us now ask a different question, namely, what is the limit of
Eq.~(\ref{eq4:4}) for $r\gg\Gamma_1$ and how is it approached? We
expand the arcustangent in Eq.~(\ref{eq4:4}) for large argument
and arrive at
\begin{equation}
\Delta\Sigma^2(r,\lambda) \approx r - \frac{2}{\pi}\Gamma_1 - 
\frac{\Gamma_1}{\pi}\ln\left(1+\left(\frac{r}{\Gamma_1}\right)^2\right)
\ .
\label{eq4:6}
\end{equation}
Apart from a constant and logarithmic corrections in $r$ (in this
context the GUE result $\Sigma^2_{\rm GUE}(r)$ to be added to
Eq.~(\ref{eq4:6}) is also nothing but a logarithmic correction) we
recover the Poisson limit $\Sigma^2_{\rm Poi}(r) = r$. The energy
scale governing the asymptotic approach to the Poisson limit is given
by $\Gamma_1$. We can see this by dividing Eq.~(\ref{eq4:4}) by $r$.
Within our usual approximations we find
\begin{equation}
\frac{1}{r}\Delta\Sigma^2(r,\lambda) = \frac{2}{\pi}{\rm
  arctan}\left(\frac{r}{\Gamma_1}\right) - 
\frac{\Gamma_1}{\pi
  r}\ln\left(1+\left(\frac{r}{\Gamma_1}\right)^2\right)   \ .
\label{eq4:7}
\end{equation}
This function approaches $1$ for $r\gg\Gamma_1$ and depends only on
the ratio $r/\Gamma_1$. Hence the relevant energy scale is $\Gamma_1$
as claimed.

It is quite instructive to compare our results with the case of a
disordered metal in the diffusive limit. The two-point level
correlator for this case was calculated by Al'tshuler and
Shklovskii\cite{Alt86} in the framework of diagrammatic perturbation
theory,
\begin{equation}
\label{eq4.7a}
X_2\left(\frac{\omega}{D}\right)=1+\frac{D^2}{\beta\pi^2}\ 
\mbox{Re}\ \sum_{\bf q}\ \frac{1}{(-i\omega+{\cal D}_0\,{\bf q}^2)^2}
\end{equation}
where $\beta=1,\ 2,\ 4$ labels the symmetry class and ${\cal D}_0$ is 
the diffusion constant (note that $D$ is the mean level spacing). 
This expression is valid for $\omega\gg D$. The sum extends over 
the diffuson modes characterized by wave vectors ${\bf q}$. The mode 
${\bf q}=0$ corresponds to random matrix theory while the modes 
${\bf q}\neq 0$ give the deviations from random matrix behavior. The latter 
can be described by a function $f(x)$ in the following way 
\begin{equation}
\label{eq4.7b}
X_2\left(\frac{\omega}{D}\right)=1-\frac{D^2}{\beta\pi^2\,\omega^2}
+\frac{D^2}{E_c^2}\ f\left(\frac{\omega^2}{E_c^2}\right),\quad
f(x)=a_0+a_2 x^2+\ldots\ .
\end{equation}
Here $E_c\sim{\cal D}_0/L^2$ is the diffusive Thouless energy for a
sample of size $L$. The coefficients $a_0, a_2, \ldots$ and therefore
the function $f(x)$ depend only on the space dimension $d$ and on the
particular geometry of the sample but not on any other physical
parameter.  These coefficients are given by discrete sums over 
``integer'' wave vectors which converge nicely for $d\le 3$.  The
correction of the nonzero modes to the number variance is immediately
estimated as
\begin{equation}
\label{eq4.7c}
\Delta\Sigma^2(r)\sim \frac{r^2}{(E_c/D)^2}+\ldots
\end{equation}
with $r=\omega/D$. This corrections is of order 1 just for $r\sim E_c/D$ 
and not $\sqrt{E_c/D}$. Here the crossover is indeed governed 
by only one parameter and there is no transition regime like the interval 
between $\sqrt{\Gamma_1}$ and $\Gamma_1$ as found above.

\subsection{Comparison with Numerical Spectra}
\label{sec4_2}

For a verification of our analytical calculations we numerically 
generate a matrix ensemble of the type (\ref{eq1}). The distribution
function $p^{(0)}(\eta)$ of the independent elements of the diagonal
matrix $H_0$ is chosen to be a semicircle with radius $R_0=30$ (in
arbitrary energy units). The matrix $H_0$ is perturbed by a GUE
ensemble $H_1$ with $\langle |H_1^{ij}|^2\rangle = \alpha^2/2$ as in
Eq.~(\ref{eq1}). We consider matrices of dimension $N=500$ and for
each set of values $\alpha, R_0$ we take $500$ realizations into
account. To ensure that $\Gamma_1 > D$ in our simulation we take the
parameter $\alpha$ from the interval $[0.1,3.0]$. Larger values of
$\alpha$ are no longer useful because it is difficult to distinguish
the ensuing spectral statistics from pure GUE statistics.

The spectra obtained by diagonalization of $H=H_0 +H_1$ are
unfolded numerically. Then the number variance $\Sigma^2_{\rm sim}(r)$
is calculated for a number of different values of $\alpha$. Given this
set of numerical functions we fit (by hand) the theoretical formula
given by Eqs.~(\ref{eq4:3}) and (\ref{eq4:4}) to the data. This
allows to determine the unfolded crossover parameter $\lambda$ (via the
relation $\Gamma_1 = \pi\lambda^2$). Furthermore, we also extract the
``critical'' energy scale $r_c$ at which $\Sigma^2_{\rm sim}(r)$
deviates from  GUE behavior from the data. We define $r_c$ by the
condition 
\begin{equation}
\frac{\Sigma^2_{\rm sim}(r_c) - \Sigma^2_{\rm GUE}(r_c)}{
  \Sigma^2_{\rm GUE}(r_c)}  = 0.2, 
\label{eq4:8}
\end{equation}
which means that at the point $r_c$ the actual (simulated or fitted)
number variance exceeds $ \Sigma^2_{\rm GUE}(r_c)$ by $20$ percent. Of
course, this condition is somewhat arbitrary. However, in our opinion
it correctly reflects what we usually mean by a ``significant
deviation from GUE behavior''.  By means of the procedure just
described a given value of $\alpha$ is associated with particular
values of $\lambda$ and $r_c$. The following figures illustrate the
procedure and our results.

\begin{figure}
\centerline{
\psfig{file=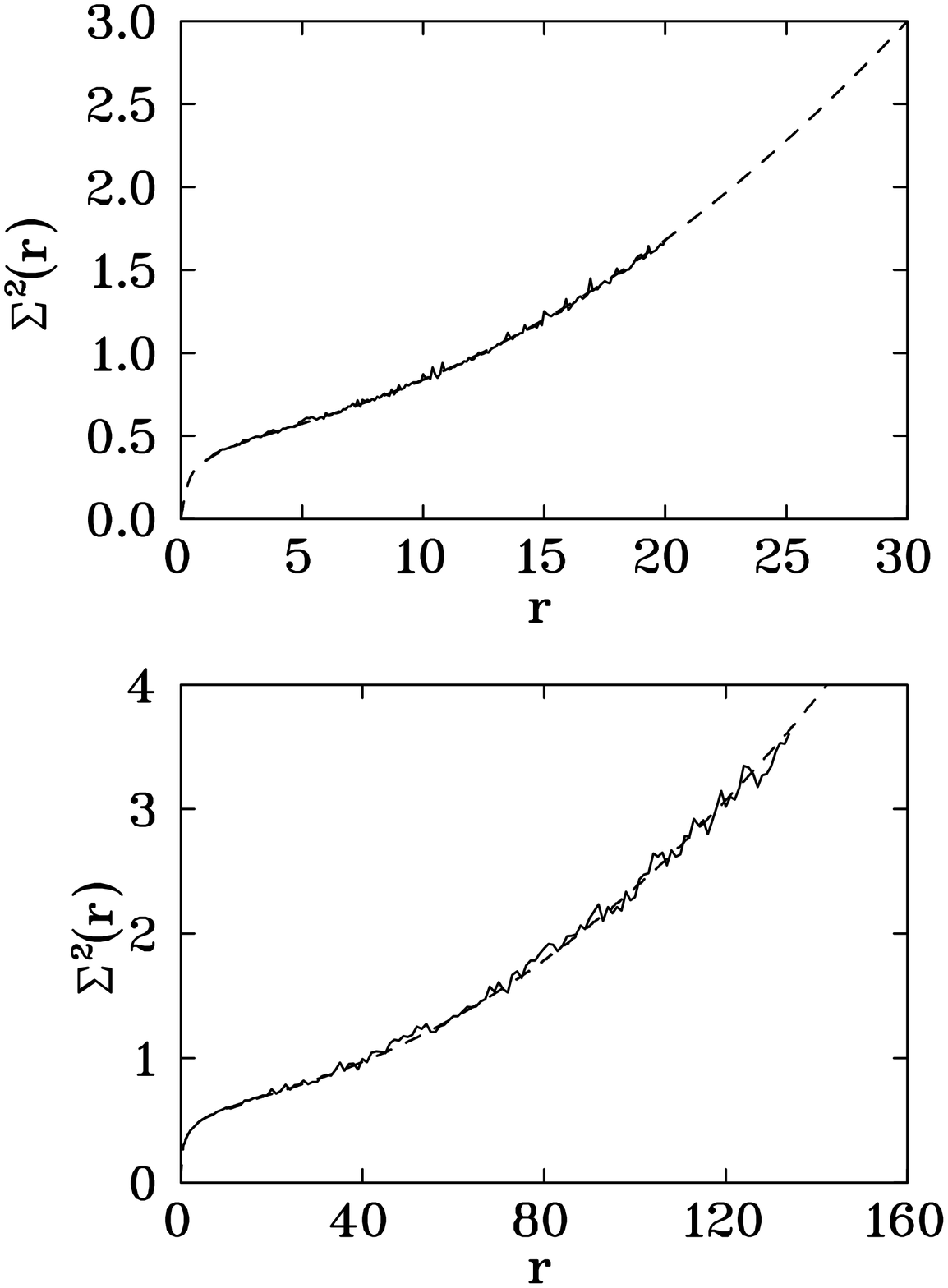,width=5.0in}
}

\vspace{0.3cm}

\caption{
The number variance $\Sigma^2(r)$ as obtained from the numerical
simulation (full line) and from our analytical calculation for
appropriately chosen $\lambda$ (dashed line). The top and the bottom
part of the figure correspond to $\alpha = 0.5$ and $\alpha = 1.2$
respectively. 
}
\label{fig2}
\end{figure}

\begin{figure}
\centerline{
\psfig{file=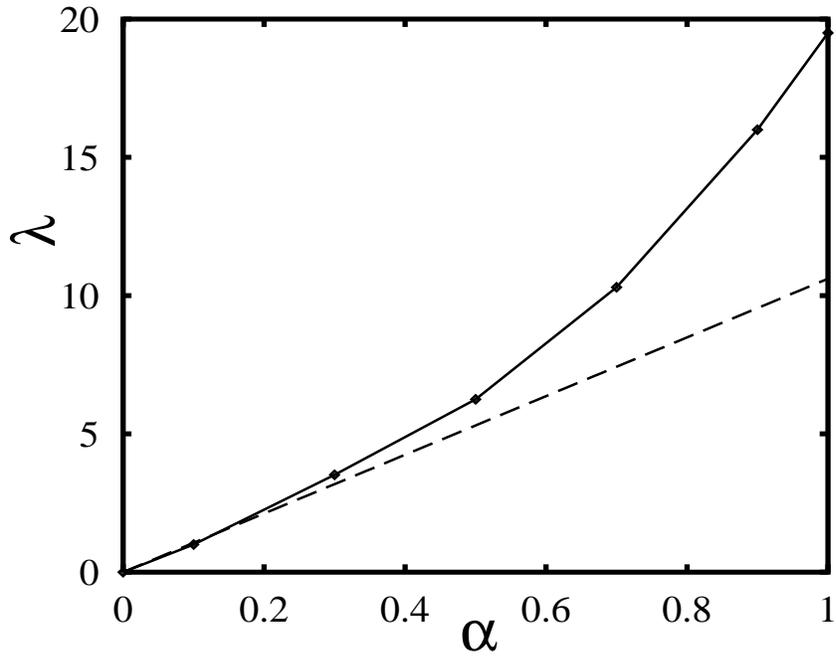,width=5.0in}
}

\vspace{0.3cm}

\caption{
The parameter $\lambda$ as a function of $\alpha$. 
Shown is the value of $\lambda$ obtained from the fit of our analytical 
formula for the number variance to the simulated data (full line and points) 
and the theoretical expression $\lambda_{\rm th}=\alpha/D$ 
(dashed lined). 
Here $D$ is the mean level spacing for the parameter values mentioned 
in the text. Note that the critical value $\alpha_c\approx 0.95$ 
corresponds to the situation in which the Breit--Wigner width 
$\Gamma_1\,D$ equals the radius $R_0$ of the unperturbed semicircle 
density. 
}
\label{fig2b}
\end{figure}

\begin{figure}
\centerline{
\psfig{file=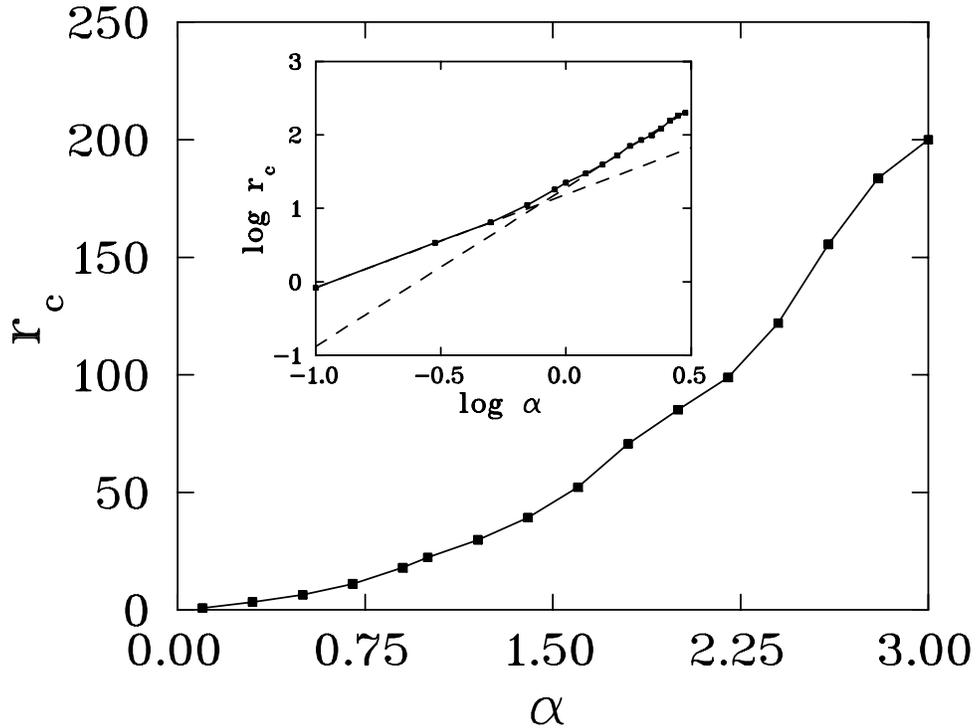,width=5.0in,angle=90}
}

\vspace{0.3cm}

\caption{
The dependence of the critical energy $r_c$ as defined by the 20
percent criterion explained in the text on the strength parameter
$\alpha$. Inset: The same on a logarithmic scale. In addition we have
drawn two dashed lines with slopes 1.2 and 2.2, respectively, to guide
the eye.
 }
\label{fig3}
\end{figure}

\begin{figure}
\centerline{
\psfig{file=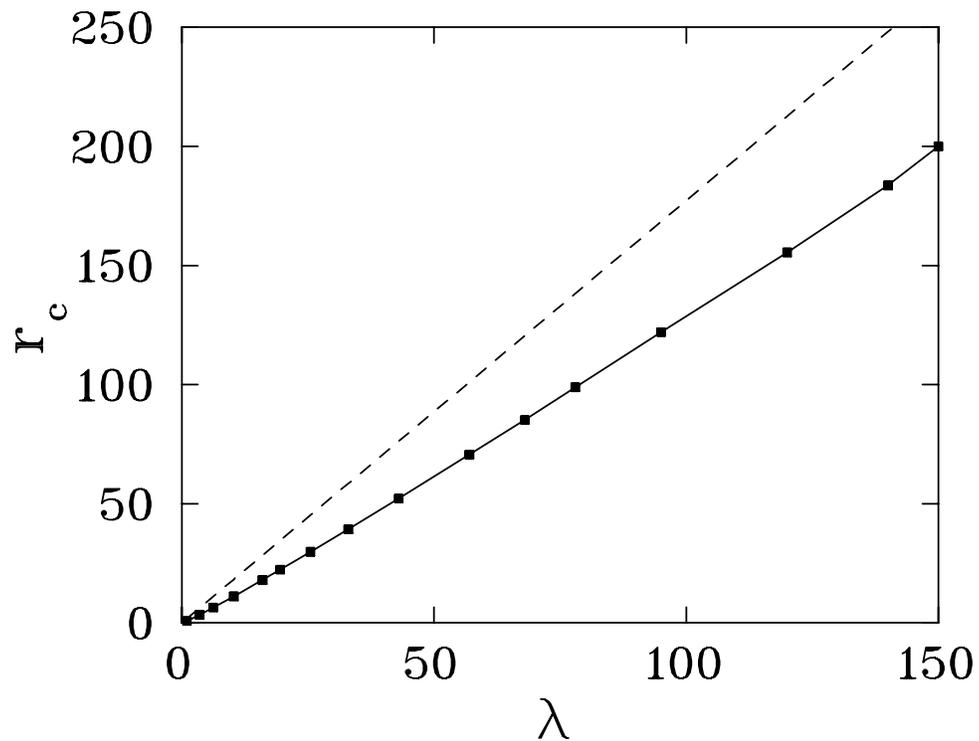,width=5.0in,angle=90}
}

\vspace{0.3cm}

\caption{
The dependence of the critical energy $r_c$ on the unfolded strength
parameter $\lambda$ (full line). The appropriate value of $\lambda$ is
obtained from a fit of our analytical formula to the simulated data. 
For comparison we also show the theoretical energy scale
$\protect\sqrt{\Gamma_1} = \protect\sqrt{\pi}\lambda$ (dashed
line). Both curves are 
roughly proportional to each other.
 }
\label{fig4}
\end{figure}

In Fig.~\ref{fig2} we show the numerically calculated number variance
$\Sigma^2_{\rm sim}(r)$ and the corresponding fit for the two values
$\alpha = 0.5$ and $\alpha = 1.2$. In the latter case, there is very
satisfactory agreement up to very large energy separations of $r=140$.
In the former case, we find agreement of this quality up to $r=30$.
Beyond this value, we find some disagreement which is likely to be due
to the limits of our approximation which relies on large $\lambda$
values (i.e. $\alpha\gg D$).  Fig.~\ref{fig2b} shows the fitted
crossover parameter $\lambda$ and the theoretical value $\lambda_{\rm
  th}=\alpha/D$.  For small values $\alpha\lesssim 0.5$ both curves
agree well.  The deviations for $\alpha\gtrsim 0.5$ are due to the
fact that the Breit--Wigner width $\Gamma_1 D$ already approaches the
effective bandwidth of the unperturbed spectrum. In fact the radius
$R_0$ of the semicircle density equals $\Gamma_1 D$ for
$\alpha_c\approx 0.95$.  In this regime both theoretical approaches
are strictly speaking not valid due to the condition (\ref{eq24}).
However, it seems that the theoretical expression for the number
variance remains even here correct, if the parameter $\lambda$ is
correspondingly rescaled (``fitted'').  Fig.~\ref{fig3} shows the
dependence of the critical energy separation $r_c$ on $\alpha$. Two
observations are important: First, the dependence is by no means
linear. Second, the data seems to indicate a crossover between two
power--laws (see inset for a logarithmic plot), which occurs around
$\alpha\approx 0.6-0.7$.  The approximate power--law exponents in the
two regions are roughly given by $1.2$ and $2.2$, respectively.
Fig.~\ref{fig4}, on the other hand, demonstrates that the dependence
of $r_c$ on $\lambda$ is almost perfectly linear as expected on
theoretical grounds and does not exhibit any crossover behavior. The
dashed line in Fig.~\ref{fig4} represents the theoretical energy scale
$\sqrt{\Gamma_1} = \sqrt{\pi}\lambda$. We see that the 20 percent
criterion gives roughly $r_c\approx 0.7 \sqrt{\Gamma_1}$.

The difference between Figs.~\ref{fig3} and \ref{fig4} is due to the
deviations shown in Fig.~\ref{fig2b} for $\alpha\gtrsim 0.5$. This
effect is related to the finite bandwidth being comparable to the
Breit--Wigner width and produces the crossover in Fig.~\ref{fig3}.  In
this context, we mention another effect which can play a role in such
numerical simulations. The mean level spacing of the matrix ensemble
(\ref{eq1}) in general depends on $\alpha$. For large enough values of
this parameter the perturbation $\alpha H_1$ in Eq.~(\ref{eq1}) will
start to increase the overall bandwidth of the spectrum and hence the
mean level spacing. Therefore the dimensionless critical energy
$r_c(\alpha) = \omega_c(\alpha)/D(\alpha)$ in general receives
contributions from both the numerator and denominator.

In Fig.~\ref{fig4}, we do not observe any crossover behavior showing
that $r_c$ is a linear function of $\lambda$. This confirms our
previous analytical result that the relevant energy scale for
deviations from the GUE is given by $\sqrt{\Gamma_1}\propto\lambda$.

\section{Summary and Conclusions}
\label{sec5}

The main results of the present paper are the exact expression for
$X_2(r)$ in terms of a handy double integral in Eq.~(\ref{eq15}), the
perturbative formula (\ref{eq51}) for $X_2(r)$ including the extra
terms (\ref{eq62}), and the perturbative expressions for $\Sigma^2(r)$
in Eqs.~(\ref{eq4:3}) and (\ref{eq4:4}). Our calculations show that
the spectral properties of the Hamiltonian $H = H_0 + \alpha H_1$ are
governed by two energy scales, the Breit--Wigner width $\Gamma_1$ and
its square root $\sqrt{\Gamma_1}$. While the former sets the scale for
deviations from the Poisson limit, the latter determines the scale at
which the difference to GUE statistics becomes noticeable.

To derive these results we employed two supersymmetric approaches, the
graded eigenvalue method and the coset method. For the technically
interested expert, the problem considered in this paper offers the
possibility to compare both methods in some detail. As a new physical
insight from this comparison, we could show that random fluctuations
of the Breit--Wigner width lead to certain extra contributions to the
spectral correlation functions.  These contributions are automatically
included in the (exact) graded eigenvalue method. They are also
included in the coset method but here the discrete saddle point
equation requires a very careful treatment when applying the
additional average with respect to the diagonal elements because the
Breit--Wigner width exhibits small random fluctuations when those
diagonal elements are varied.

We briefly discuss the relevance and implications of this paper for
previously published work. In Ref.~\onlinecite{Alt97} a diagrammatic
analysis of the two--level correlation function $X_2(r)$ was
presented. There, a central aim was to provide a physically more
transparent derivation of the results published in
Ref.~\onlinecite{Guh97}. Comparing our formula (\ref{eq20}) with the
expression (15) in Ref.~\onlinecite{Alt97} it seems that the results
of Ref.~\onlinecite{Alt97} are qualitatively, but not quantitatively,
correct.  This discrepancy, which appears for the second order terms,
is due to the contributions arising from the statistical fluctuations
of the Breit--Wigner width (see Sec.~\ref{sec3_5} and
Appendix~\ref{app3}) which are completely neglected in
Ref.~\onlinecite{Alt97}. To understand this in detail, we have also to
take into account that Altland {\it et al.} considered a model where
the diagonal matrix elements of the perturbation $\alpha H_1$ are
zero. We have therefore modified their diagrammatic calculation for
our case where $H_1$ is a GUE-matrix with non-vanishing diagonal matrix
elements. The result is in perfect agreement with Eq.~(\ref{eq51})
obtained by the $\sigma$ model perturbation theory. Adding the extra
terms (\ref{eq61}) from the fluctuating Breit--Wigner width, we again
recover our result (\ref{eq20}) from the graded eigenvalue method.

However, we expect that the model of Altland et al. (with
$H_1^{jj}=0$) should be equivalent to our case since, in the limit
$D\,\Gamma_1/B\to 0$, the diagonal elements can be absorbed by an
infinitesimal rescaling of $H_0$. Inspecting Eq.~(15) of
Ref.~\onlinecite{Alt97}, we see that here the diffuson perturbation
theory already contains the first extra contribution (\ref{eq_c12})
but not the second extra term (\ref{eq_c25}). The reason is that for
the model with $H_1^{jj}=0$ the fluctuations of $\Gamma$ appear in a
different way. Indeed, the Dyson (or saddle point) equation is now
slightly modified
\begin{equation}
\label{eq4.9_neu}
\Gamma(k)=-\frac{\Gamma_1 D}{\pi}\sum_{j(\neq k)}
\frac{1}{x-H_0^{jj}+D\Gamma(j)/2}\ .
\end{equation}
Note that $\Gamma(k)$ depends on the site index and that the sum
extends only over the $j$ being different from $k$. In the
approximation (\ref{eq32}) this gives no difference while the
treatment of Appendix \ref{app3} is modified.  It turns out that now
the first term (\ref{eq_c12}) vanishes while the second contribution
(\ref{eq_c25}) still persists. Together with the contributions from
the diffuson diagrams [Eq.~(15) of Ref.~\onlinecite{Alt97}], we obtain
again perfect agreement with Eq.~(\ref{eq20}).

Apparently, for both models with and without diagonal matrix elements
in $H_1$ there are second order contributions arising from both the
diagrammatic diffuson terms and the fluctuations of the Breit-Wigner
width. The sum of these contributions is in both cases the same as it
should be. However, the way they appear as diffuson terms or
$\Gamma$-fluctuations is different.

In Ref.~\onlinecite{Wei96} numerical and analytical evidence was
reported for a crossover from a linear to a quadratic dependence of
the critical energy $r_c$ on the strength of the perturbing (GOE)
matrix elements.  From our calculations we cannot confirm such a
behavior. Both our previous~\cite{Guh97} and present analysis indicate
the the critical energy $r_c$ is linear in $\lambda$, irrespective of
the range, $\lambda>1$ or $\lambda<1$ (see Ref.~\onlinecite{Guh97}),
considered. Fig.~\ref{fig3} gives the (erroneous) impression of a
crossover because the strength parameter $\alpha$ is not properly
measured on the unfolded scale. The fact that the mean level spacing
of the Hamiltonian (\ref{eq1}) depends on $\alpha$ necessitates an
$\alpha$--dependent unfolding. This leads to the linear behavior in
Fig.~\ref{fig4}.

Finally, we would like to re-emphasize that the ensemble (\ref{eq1})
simulates a very general situation: A diagonalized quantum system with
Poisson statistics is perturbed by a non-diagonal random operator.
Situations where this model system is appropriate should abound, and
in these cases the spectral behavior predicted here, in particular the
energy scales $\sqrt{\Gamma_1}$ and $\Gamma_1$ should be readily
identifiable.

\section*{Acknowledgment}

It is a pleasure to thank Shinji Iida for his interest in our work and
very valuable and helpful discussions and remarks.

\appendix

\section{Perturbative evaluation of Eq.~(\protect\ref{eq17})}
\label{app1}

In this Appendix we compute the terms of order $c^1$ and $c^2$ 
in the perturbative expansions of the integral $P(c,\kappa)$ 
defined in Eq.~(\ref{eq17}). For this we keep in both sums in
Eq.~(\ref{eq17})  the terms up to order $\rho^2$,
\begin{equation}
\label{eq_a1}
P(c,\kappa)=\frac{1}{2\pi}\int_0^{2\pi} d\varphi\int_0^\infty
d\rho\ \rho\,e^{i\varphi}\,\cos\varphi\left(1-\frac{\rho}{\kappa}
\sin\varphi\right)
\left(1+\frac{\rho\,e^{i\varphi}}{2\kappa\,i}+
\Big(\frac{\rho\,e^{i\varphi}}{2\kappa\,i}\Big)^2+\cdots\right)
\ e^{-L_0-L_1-L_2-\cdots}
\end{equation}
with
\begin{equation}
\label{eq_a2}
L_0=\frac{\rho^2}{2\Pi}\quad,\quad \Pi=\frac{ck}{k+i}\quad,
\quad L_n=\frac{i\rho^2}{2c\kappa}\left(\frac{\rho\sin\varphi}{\kappa}
\right)^n\quad \mbox{if}\ n\ge 1\quad.
\end{equation}
Here we have introduced the notion $\Pi$ for the ``propagator'' of the 
perturbation theory. To keep a compact notation we introduce the abbreviation
\begin{equation}
\label{eq_a3}
\langle f(\rho,\varphi)\rangle_{(\rho,\varphi)}=
\frac{1}{2\pi}\int_0^{2\pi} d\varphi\int_0^\infty d\rho\ \rho\ 
e^{-\rho^2/(2\Pi)}\ f(\rho,\varphi) \ ,
\end{equation}
where $f(\rho,\varphi)$ stands for an arbitrary function of $\rho$ and 
$\varphi$. Expanding the exponential in (\ref{eq_a1}) with respect 
to $L_1$ and $L_2$, we obtain
\begin{equation}
\label{eq_a4}
P(c,\kappa)=\left\langle e^{i\varphi}\,\cos\varphi
\left(1+\frac{\rho\,e^{-i\varphi}}{2\kappa i}+
\Big(\frac{\rho}{2\kappa i}\Big)^2+\cdots\right)
\left(1-L_1-L_2+\frac{1}{2} L_1^2+\cdots\right)
\right\rangle_{(\rho,\varphi)}\quad.
\end{equation}
Obviously, in (\ref{eq_a4}) the terms with odd power in $\rho$ vanish 
while the terms with even power can be evaluated using 
$\langle \rho^{2m}\rangle_{(\rho,\varphi)}=\Pi^{m+1}\,2^m\,m!$.
The $c^1$-term is therefore given by
\begin{equation}
\label{eq_a5}
P^{(1)}(c,\kappa)=\langle e^{i\varphi}\cos\varphi\rangle_{(\rho,\varphi)}=
\frac{1}{2}\,\Pi
\end{equation}
in agreement with Eq. (\ref{eq18}). There are four different 
second order terms,
\begin{eqnarray}
\label{eq_a6}
P_a^{(2)}(c,\kappa)&=&\left\langle e^{i\varphi}\cos\varphi
\Big(\frac{\rho}{2\kappa i}\Big)^2
\right\rangle_{(\rho,\varphi)}
=-\frac{1}{4\kappa^2}\langle \rho^2\cos^2\varphi\rangle_{(\rho,\varphi)}
=-\frac{\Pi^2}{4\kappa^2}\ ,\\
\label{eq_a7}
P_b^{(2)}(c,\kappa)&=&-\left\langle e^{i\varphi}\cos\varphi\,
\frac{\rho\, e^{-i\varphi}}{2\kappa i}\,L_1\right\rangle_{(\rho,\varphi)}=
-\frac{1}{4c\kappa^3}\langle \rho^4\cos\varphi\,\sin\varphi
\rangle_{(\rho,\varphi)}=0\ ,\\
\label{eq_a8}
P_c^{(2)}(c,\kappa)&=&-\langle e^{i\varphi}\cos\varphi\,L_2\rangle=
-\frac{i}{2c\kappa^3}\,\langle \rho^4\cos^2\varphi\,\sin^2\varphi\rangle
=-\frac{i}{2c\kappa^3}\,2!\cdot2^2\,\Pi^3\frac{1}{8}
=-\frac{i}{2c\kappa^3}\,\Pi^3\ ,\\
\label{eq_a9}
P_d^{(2)}(c,\kappa)&=&\frac{1}{2}\langle e^{i\varphi}\cos\varphi\,L_1^2
\rangle=-\frac{1}{8c^2\kappa^4}\langle \rho^6 \cos^2\varphi\,
\sin^2\varphi\rangle=
-\frac{1}{8 c^2\,\kappa^4}\,3!\cdot 2^3\,\Pi^4\,\frac{1}{8}=
-\frac{3}{4 c^2\,\kappa^4}\,\Pi^4\ ,
\end{eqnarray}
which result in
\begin{equation}
\label{eq_a10}
P^{(2)}(c,\kappa)=\sum_{\alpha=a,b,c,d} P_\alpha^{(2)}(c,\kappa)
=-\frac{\Pi^2}{4\kappa^2}\,\left(1+2i\Big(\frac{\Pi}{c\kappa}\Big)
+3\Big(\frac{\Pi}{c\kappa}\Big)^2\right)\ .
\end{equation} 
Inserting $\Pi=c\kappa/(\kappa+i)$, we obtain Eq.~(\ref{eq20}). 

\section{Pertubation theory in the frame work of the coset method}
\label{app2}

The precise $\sigma$ model action ${\cal L}_0$ and the preexponential 
terms in (\ref{eq40}) still depend on the particular realisation of 
the diagonal matrix elements $H_0^{jj}$. The aim of this appendix is to 
show how one can perform in a suitable way the average with respect to these 
matrix elements and how the resulting $Q$-integral can be evaluated 
perturbatively for $Q$ being close the origin $\Lambda$ in the coset space. 
In order to expand ${\cal L}_0$ with respect to 
$\Delta Q=\{\Lambda,Q\}/2-1=2R^2$, we consider the quantity 
\begin{equation}
\label{eq_b1}
I=\mbox{trg}\ \ln(a+b\Lambda+c Q)
\end{equation}
with some complex parameters $a,\ b,\ c$ and $\Lambda$, $Q$ as defined 
in section \ref{sec3_3}. 
The quantity $I$ only depends on the radial parameters \cite{Efe83} $\theta$ 
of $Q=T^{-1}\Lambda T$, i.e. we may choose $T=\exp(\theta\sigma_1)$, where 
$\sigma_1={0\ 1\choose 1\ 0}$ and $\theta$ is $2\times 2$ supermatrix with 
two independent radial parameters \cite{Efe83} (for the unitary 
symmetry class). Then we have $\sigma_1\Lambda\sigma_1=-\Lambda$ and 
$\sigma_1 Q\sigma_1=-Q$. Inserting $1=\sigma_1^2$ and permuting the matrix 
products in the argument of the logarithm in (\ref{eq_b1}), we thus obtain:
\begin{eqnarray}
I & = & \mbox{trg}\ \ln(a-b\Lambda-c Q)=
\frac{1}{2}\mbox{trg}\ \ln\Bigl((a+b\Lambda+c Q)(a-b\Lambda-c Q)\Bigr)=
\nonumber\\
\label{eq_b2}
& = & \frac{1}{2}\mbox{trg}\ \ln\biggl(\Bigl[a^2-(b+c)^2\Bigr]-bc
\Bigl[Q\Lambda+\Lambda Q-2\Bigr]\biggr)=
\frac{1}{2}\mbox{trg}\ \ln\biggl(\Bigl[a^2-(b+c)^2\Bigr]-2bc\, \Delta Q
\biggr)\\
\nonumber
& = & \frac{1}{2}\mbox{trg}\ \ln\biggl(1-\frac{2bc}{a^2-(b+c)^2}\ \Delta Q
\biggr)\quad.
\end{eqnarray}
Using this identy, we can rewrite ${\cal L}_0$ as 
\begin{equation}
\label{eq_b3}
{\cal L}_0=\frac{1}{2}\sum_j \mbox{trg}\ \ln
\left(1-\frac{i\,\omega\Gamma_1 D/2}{(x-H_0^{jj}+\Gamma_0 D/2)^2
+(\Gamma_1 D/2-i\omega/2)^2}\ \Delta Q\right)\quad.
\end{equation}
The logarithm can be exanded in powers of $\Delta Q$ giving 
${\cal L}_0={\cal L}_0^{(2)}+{\cal L}_0^{(4)}+\cdots$ where 
${\cal L}_0^{(2)}$ is the gaussian part of the action. Furthermore 
we note that in the limit $\Gamma_1\gg 1$ the 
discrete $j$-sum consists of Lorentzians with 
well overlapping resonances which allows to replace in lowest order 
of $\Gamma_1^{-1}$ the gaussian part ${\cal L}_0^{(2)}$ by its average 
over the diagonal elements $H_0^{jj}$. Therefore we obtain, 
in the limit $B/\Gamma_1\to \infty$, 
\begin{equation}
\label{eq_b4}
{\cal L}_0^{(2)}(Q)\approx [ {\cal L}_0^{(2)}(Q)]_{0}
=-i\,\frac{\pi}{D}\,\frac{\omega\Gamma_1/2}{\Gamma_1 D-i\omega}
\,\mbox{trg}(\Delta Q)
=-i\,\frac{\pi}{2}\,\frac{r\Gamma_1}{\Gamma_1-ir}\,\mbox{trg}(\Delta Q)
\end{equation}
where $r=\omega/D$ and $B$ is the bandwidth over which the diagonal matrix 
elements $H_0^{jj}$ are distributed. Omitting the details, we mention 
that one can show in a more refined way that the statistical 
fluctuations indeed yield corrections of lower order in $\Gamma_1^{-1}$. 
For this, one has to analyze the average of the quantity 
$\exp(-{\cal L}_0^{(2)}(Q))$ more carefully. 

The spirit of the pertubative calculation consists of expanding the 
the preexponential terms [and also of the nongaussian terms 
$\exp(-{\cal L}_0^{(4)}-\cdots)$] with respect to the variables 
$R$ (or $t$) [see (\ref{eq45})]. Then one 
can systematically evaluate the gaussian integrals using Wick's theorem. 
Each pair of $t$ and $\bar t$ yields here a power $\Pi_D^{-1}$ of the 
propagator which is defined by the prefactor of the gaussian action 
\begin{equation}
\label{eq_b5}
{\cal L}_0^{(2)}(Q)\approx 
-\pi\,i\,\frac{r\Gamma_1}{\Gamma_1-ir}\,\mbox{trg}(R^2)
=-2\pi\,i\,\frac{r\Gamma_1}{\Gamma_1-ir}\,\mbox{trg}(\bar t\,t)
=\Pi_D\,\mbox{trg}(\bar t\,t)\quad.
\end{equation}
The perturbation theory is justified for $|\Pi_D|\gg 1$ which is 
certainly correct if $\Gamma_1\gg 1$ and if the ratio $\kappa=r/\Gamma_1$ 
is finite and non $0$. 
As in the pure GUE case, this simple perturbation theory 
does not allow to access the very small frequencies $\omega\sim D$. 
We emphasize that this is different for the pertubative evaluation 
of the integral (\ref{eq17}) which was obtained in the frame work 
of the precise graded eigenvalue method in section \ref{sec2} (see 
appendix \ref{app1}). Here the pertubative condition only consists of 
$\Gamma_1\gg 1$ and the behavior for small $\omega$ is precisely 
described by the residuum contributions in (\ref{eq14}) which even contain 
the nonanalytical corrections. 

We have to expand $g_j$ in the preexponential factors in (\ref{eq40}) 
with respect to the small quantity
\begin{equation}
\label{eq_b6}
\delta Q=Q-\Lambda=2\Lambda R\sqrt{1+R^2}+2\Lambda R^2=
2\Lambda R+2\Lambda R^2+\cdots
\end{equation}
which gives
\begin{equation}
\label{eq_b7}
g_j=g_j^{(0)}+\left(i\frac{\Gamma_1 D}{2}\right)
\ g_j^{(0)}\,\delta Q\,g_j^{(0)}+
\left(i\frac{\Gamma_1 D}{2}\right)^2\ g_j^{(0)}\,\delta Q\,g_j^{(0)}
\,\delta Q\,g_j^{(0)}+\cdots
\end{equation}
where 
\begin{eqnarray}
\label{eq_b8}
g_j^{(0)}&=&\left[x-H_0^{jj}+\Gamma_0\,D/2
+i\Lambda\,(\Gamma_1\,D-i\omega)/2\right]^{-1}\quad,\\
\label{eq_b9}
\bar g_j^{(0)}&=&\left[x-H_0^{jj}+\Gamma_0\,D/2
-i\Lambda\,(\Gamma_1\,D-i\omega)/2\right]^{-1}\quad.
\end{eqnarray}
To evaluate the term $S_1(Q)$ in (\ref{eq40}), we consider 
\begin{eqnarray}
\label{eq_b10}
\sum_j\mbox{trg}(g_j\,P_{\pm})&=&
\sum_j\mbox{trg}(g_j^{(0)}\,P_{\pm})
+\left(i\frac{\Gamma_1 D}{2}\right)\sum_j
\mbox{trg}[(g_j^{(0)})^2\,(2\Lambda R+2\Lambda R^2)\,P_{\pm}]\\
\nonumber
&&+\left(i\frac{\Gamma_1 D}{2}\right)^2\sum_j\mbox{trg}
[(g_j^{(0)})^2\,\bar g_j^{(0)}\,(2\Lambda R\,2\Lambda R)\,P_{\pm}]+\cdots\\
\nonumber
&=&:A_0^{(\pm)}+A_1^{(\pm)}+A_2^{(\pm)}+
\end{eqnarray}
where we have used that $g_j^{(0)}\,R=R\,\bar g_j^{(0)}$ since
$\Lambda R\Lambda =-R$. The term $S_1(Q)$ is now given by
\begin{equation}
\label{eq_b11}
S_1(Q)=\left(A_0^{(+)}+A_1^{(+)}+A_2^{(+)}\right)
\left(A_0^{(-)}+A_1^{(-)}+A_2^{(-)}\right)
\end{equation}
The main contribution arises from the product 
$A_0^{(+)}\,A_0^{(-)}$ which is effectively of order $\Pi_D^{-0}$ 
and also contains terms of higher orders 
$\Pi_D^{-1},\ \Pi_D^{-2},\ldots$. This contribution 
just corresponds to the boundary terms (\ref{eq41}) which are extensively 
discussed in subsection \ref{sec3_5} and appendix \ref{app3}. 
Here we concentrate on the higher order terms arising from $A_1$ and $A_2$ 
which are both of second order in $R$ (the linear $R$ term in $A_1$ 
vanishes under the graded trace). They will give contributions of 
order $\Pi_D^{-2}$ which is the highest order in which we are interested. 
(This corresponds to order $c^2$ in appendix \ref{app1}.) 
Therefore we can here replace the correlated $H_0$-averages 
like $[A_2\, A_2\,\exp(-{\cal L}_0^{(2)})]_0$ by 
$[A_2]_0\,[A_2]_0\,\exp(-[{\cal L}_0^{(2)}]_0)$ since the correlations 
arising from the diagonal terms or the exponential are already of order 
$\Pi_D^{-3}$. 
The crossed terms like $[A_0\, A_2]_0$ are in principle of order 
$\Pi_D^{-1}$ but they vanish upon the gaussian contraction rules 
(for details see Ref. \onlinecite{Iid90}). Actually, they are of the type 
$\mbox{trg}(t\,\bar t\,P_+)\to \Pi_D^{-1}\mbox{trg}(1)\,\mbox{trg}(P_+)=0$. 
The product of these terms with first term arising form the nongaussian 
part of the action, i.e. $[A_0\, A_2]_0\,{\cal L}_0^{(4)}$, is of order 
$\Pi_D^{-2}$ and of the type 
$\mbox{trg}(t\,\bar t\,P_+)\,\mbox{trg}(t\,\bar t\,t\,\bar t)$ 
which also vanishes. The term $[A_1]_0\,[A_1]_0$ vanishs upon 
the $H_0$-average while $[A_2]_0\,[A_2]_0$ gives the only 
nonvanishing contribution. The $H_0$-average can be performed 
using the integrals (\ref{eq_c6},\ref{eq_c8}) evaluated in appendix \ref{app3}:
\begin{equation}
\label{eq_b12}
A_2^{(\pm)}\to [A_2^{(\pm)}]_0
=\left(i\frac{\Gamma_1 D}{2}\right)^2\,\frac{(-2\pi i)}{D}\,
\frac{1}{(\Gamma_1 D-i\omega)^2}\,\mbox{trg}(-4\Lambda R^2\,P_\pm)
=-\frac{2\pi i}{D}\,\left(\frac{\Gamma_1}{\Gamma_1-i r}\right)^2
\ \left\{{ \mbox{trg}(t\,\bar t\,P_+) \atop -\mbox{trg}(\bar t\,t\,P_-) 
}\right\}\ .
\end{equation}
In summary, we obtain for the term $S_1(Q)$ 
\begin{equation}
\label{eq_b13}
[S_1(Q)]_0\approx S_1(\Lambda)+
\frac{4\pi^2}{D^2}\ \left(\frac{\Gamma_1}{\Gamma_1-i r}\right)^4
\ \mbox{trg}(t\,\bar t\,P_+)\,\mbox{trg}(\bar t\,t\,P_-) 
\end{equation}
which is just the first part of Eq. (\ref{eq47}). 

Concerning the term $S_2(Q)$, only the linear $\delta Q$-term in 
(\ref{eq_b7}) contributes
\begin{eqnarray}
\label{eq_b14}
S_2(Q)&=&\left(i\frac{\Gamma_1 D}{2}\right)^2\,
\sum_j\mbox{trg}\,[g_j^{(0)}\,(2\Lambda R)\,g_j^{(0)}\,P_+\,
g_j^{(0)}\,(2\Lambda R)\,g_j^{(0)}\,P_-]\\
\nonumber
&=&\left(i\frac{\Gamma_1 D}{2}\right)^2\,
\sum_j\mbox{trg}\,[(g_j^{(0)})^2\,(\bar g_j^{(0)})^2\,(-4)R P_+ R P_-]\ .
\end{eqnarray}
We can again perform the $H_0$-average by using (\ref{eq_c7},\ref{eq_c8}), 
\begin{equation}
\label{eq_b15}
[S_2(Q)]_0\approx \left(i\frac{\Gamma_1 D}{2}\right)^2\,(-4)\,
\frac{1}{D}\,\frac{4\pi}{(\Gamma_1 D-i\omega)^3}\,
\mbox{trg}\,[R P_+ R P_-]=
\frac{4\pi^2}{D^2}\ \frac{\Gamma_1^2}{(\Gamma_1-ir)^3}\,
\mbox{trg}[t\, P_+\,\bar t\,P_-]\ .
\end{equation}
This expression provides the second part of Eq. (\ref{eq47}). 

The gaussian contractions for (\ref{eq_b13}) and (\ref{eq_b15}) can 
be performed by
\begin{equation}
\label{eq_b16}
\mbox{trg}(t\,\bar t\,P_+)\,\mbox{trg}(\bar t\,t\,P_-) \to
\Pi_D^{-1}\mbox{trg}[\bar t\, P_+\, t\,P_-]\to
\Pi_D^{-2}\mbox{trg}(P_+)\,\mbox{trg}(P_-)=\Pi_D^{-2}\ ,
\end{equation}
and 
\begin{equation}
\label{eq_b17}
\mbox{trg}[t\, P_+\,\bar t\,P_-]\to \Pi_D^{-1}
\mbox{trg}(P_+)\,\mbox{trg}(P_-)=\Pi_D^{-1}\ ,
\end{equation}
since $\mbox{trg}(P_{\pm})=1$. Combining this with (\ref{eq_b13}) and 
(\ref{eq_b15}), we finally obtain the result (\ref{eq48}). 

We mention that this perturbative calculation can be immediately 
generalized to the orthogonal ($\beta=1$) and symplectic symmetry classes  
($\beta=4$). In these cases $t$ is a $4\times 4$-matrix containing 
$2\times 2$-blocks $t_D$ and $t_C$ for the diffuson and cooperon 
contributions. The result of this generalization just gives 
the prefactor $2/\beta$ in (\ref{eq48}).

\section{Contributions from the fluctuating Breit--Wigner width}
\label{app3}

The statistical fluctuations of the Breit--Wigner width $\Gamma$ 
discussed in subsection \ref{sec3_5} result in further second 
order contributions when the boundary term $S_1(\Lambda)$ is averaged 
over the diagonal matrix elements. To extract these contributions, 
we need the following type of simple integrals
\begin{equation}
\label{eq_c1}
I_{n,m}(\gamma)=\lim_{L\to\infty} \int_{-L}^L 
dy\ \frac{1}{(y+i\gamma/2)^n}\,\frac{1}{(y-i\gamma/2)^m}\ .
\end{equation}
Obviously, we have
\begin{equation}
\label{eq_c2}
I_{1,0}=-I_{0,1}=-i\pi\quad,\quad I_{n,0}=I_{0,n}=0\quad 
\mbox{if}\quad n\ge 2\ ,
\end{equation}
because for $n\ge 2$ the integration contour can be closed such that 
there are no singularities inside the contour. For $m=1$, the 
pol of first order inside the contour results in
\begin{equation}
\label{eq_c3}
I_{n,1}(\gamma)=(-i)^{n-1}\,\frac{2\pi}{\gamma^n}\quad,\quad n\ge 1\ .
\end{equation}
Furthermore, we obtain by partial integration the recurrence relation 
\begin{equation}
\label{eq_c4}
I_{n,m}(\gamma)=-\frac{n}{m-1}\ I_{n+1,m-1}(\gamma)
\end{equation}
which can be solved in terms of the binomial coefficient
\begin{equation}
\label{eq_c5}
I_{n,m}(\gamma)=(-1)^{n-1}\,i^{n+m-2}\left({n+m-2 \atop m-1}\right)\,
\frac{2\pi}{\gamma^{n+m-2}}\quad,\quad n,m\ge 1\ .
\end{equation}
For the subsequent calculation, we need the cases
\begin{eqnarray}
\label{eq_c6}
I_{1,1}(\gamma)&=&\frac{2\pi}{\gamma}\quad,\quad I_{1,2}(\gamma)=
-I_{2,1}(\gamma)=i\frac{2\pi}{\gamma^2}\quad,\\
\label{eq_c7}
I_{1,3}(\gamma)&=&I_{3,1}(\gamma)=-\frac{2\pi}{\gamma^3}\quad,
\quad I_{2,2}(\gamma)=2\,\frac{2\pi}{\gamma^3}\ .
\end{eqnarray}
The moments of the quantities $G_j$, $\bar G_j$ [see Eq. (\ref{eq58})] 
with respect to 
the $H_0$ average can be expressed in terms of these integrals,  
\begin{equation}
\label{eq_c8}
\sum_j[G_j^n\,\bar G_j^m]=\frac{1}{D}\lim_{B\to\infty}
\int_{-B/2}^{B/2} dH\ \frac{1}{[(x-H)+i(D\Gamma_1-i\omega)/2]^n}\,
\frac{1}{[(x-H)-i(D\Gamma_1-i\omega)/2]^m}=\frac{1}{D}\,
I_{n,m}(D\Gamma_1-i\omega)\ .
\end{equation}
Now, we turn to the average of the boundary term (\ref{eq60}) which 
we rewrite as
\begin{equation}
\label{eq_c9}
S_1(\Lambda)=(B_0+B_1)(\bar B_0+\bar B_1)
\end{equation}
with $B_0=\sum_j G_j$, $B_1=-D\,\delta\Gamma/2\sum_j G_j^2$ and analog 
for $\bar B_0$, $\bar B_1$. The main contribution $[B_0\,\bar B_0]_0$ 
has already been considered and evaluated in (\ref{eq42}). The next 
contribution is given by [see (\ref{eq57})] 
\begin{eqnarray}
\label{eq_c10}
[B_1\,\bar B_0]_0&=&-\frac{D}{2}\sum_{j,k}[\delta\Gamma\,G_j^2\,\bar G_k]_0
=\frac{D\,\Gamma_1}{2}\sum_{j,k} \left[G_j^2\,\bar G_k\left(
\frac{D}{\pi}\sum_l G_l+i\right)\right]_0\\
\nonumber
&=&\frac{D\,\Gamma_1}{2}\sum_{j} \left[G_j^2\,\bar G_j\left(
\frac{D}{\pi}\sum_l G_l+i\right)\right]_0\ .
\end{eqnarray}
The terms with $k\neq j$ vanish since all averages of the type 
$[G_j^n]_0$, $[\bar G_j^n]_0$ vanish if $n\ge 2$ [see (\ref{eq_c2})].
Furthermore, the contributions with $l\neq j$ cancel for $N\to\infty$ 
the second term because 
$\sum_l [G_l]_0=I_{1,0}(D\Gamma_1-i\omega)/D=-i\pi/D$. Therefore, 
by (\ref{eq_c7},\ref{eq_c8}) we obtain
\begin{equation}
\label{eq_c11}
[B_1\,\bar B_0]_0=\frac{D^2\Gamma_1}{2\pi}\sum_j\left[
G_j^3\bar G_j\right]_0=\frac{D^2\Gamma_1}{2\pi}\,\frac{1}{D}
\left(-\frac{2\pi}{(D\Gamma_1-i\omega)^3}\right)
=-\frac{1}{D^2}\,\frac{\Gamma_1}{(\Gamma_1-ir)^3}\ .
\end{equation}
In an analog way, one finds that $[B_0\,\bar B_1]_0$ gives the same 
contribution, such that 
\begin{equation}
\label{eq_c12}
[B_1\,\bar B_0]_0+[B_0\,\bar B_1]_0=-\frac{2}{D^2}\,
\frac{\Gamma_1}{(\Gamma_1-ir)^3}\ .
\end{equation}
The remaining contribution, 
\begin{equation}
\label{eq_c13}
[B_1\,\bar B_1]_0=\frac{D^2\Gamma_1^2}{4}\sum_{j,k}
\left[G_j^2\,\bar G_k^2\left(\frac{D}{\pi}\sum_l G_l+i\right)
\left(\frac{D}{\pi}\sum_m \bar G_m-i\right)\right]_0\equiv
C_4+C_5+C_6\quad,
\end{equation}
leads to three different subcases with 4, 5, or 6 $G$-factors. In 
the first case only the terms $j=k$ contribute
\begin{equation}
\label{eq_c14}
C_4=\frac{D^2\Gamma_1^2}{4}\sum_{j,k}\left[G_j^2\,\bar G_k^2\right]_0
=\frac{D^2\Gamma_1^2}{4}\sum_j\left[G_j^2\,\bar G_j^2\right]_0\ .
\end{equation}
For $C_5$ and $C_6$, we have to consider all different cases 
of equal or different summations indicies. In the following, we 
mention only those cases giving a nonvanishing contribution. 
Actually, for 
\begin{equation}
\label{eq_c15}
C_5=i\,\frac{D^3\Gamma_1^2}{4\pi}\sum_{j,k,m}
\left[G_j^2\,\bar G_k^2\,(\bar G_m-G_m)\right]_0
\end{equation}
there are two relevant cases:
\begin{eqnarray}
\label{eq_c16}
C_5:&\ & j=k\neq m\quad\to\quad
-2\,\frac{D^2\Gamma_1^2}{4}\sum_j\left[G_j^2\,\bar G_j^2\right]_0\ ,\\
\label{eq_c17}
C_5:&\ & j=k=m\quad\to\quad
i\,\frac{D^3\Gamma_1^2}{4\pi}\sum_j 
\left[G_j^2\,\bar G_j^2\,(\bar G_j-G_j)\right]_0\ .
\end{eqnarray}
The last term
\begin{equation}
\label{eq_c18}
C_6=\frac{D^4\Gamma_1^2}{4\pi^2}
\sum_{j,k,m,l} \left[G_j^2\,\bar G_k^2\,G_l\,\bar G_m\right]_0
\end{equation}
leads to six cases:
\begin{eqnarray}
\label{eq_c19}
C_6:&\ &l\neq(j=k)\neq m\neq l\quad\to\quad
\frac{D^2\Gamma_1^2}{4}\sum_j\left[G_j^2\,\bar G_j^2\right]_0\\
\label{eq_c20}
C_6:&\ &(j=k)\neq (m=l)\quad\to\quad
\frac{D^4\Gamma_1^2}{4\pi^2}\sum_j \left[G_j^2\,\bar G_j^2\right]_0
\sum_m\left[G_m\,\bar G_m\right]_0\\
\label{eq_c21}
C_6:&\ &(j=m)\neq (k=l)\quad\to\quad
\frac{D^4\Gamma_1^2}{4\pi^2}\sum_j \left[G_j^2\,\bar G_j\right]_0
\sum_k\left[G_k\,\bar G_k^2\right]_0\\
\label{eq_c22}
C_6:&\ &(j=k=m)\neq l\quad\to\quad
-i\,\frac{D^3\Gamma_1^2}{4\pi}\sum_j \left[G_j^2\,\bar G_j^3\right]_0\\
\label{eq_c23}
C_6:&\ &(j=k=l)\neq m\quad\to\quad
i\,\frac{D^3\Gamma_1^2}{4\pi}\sum_j \left[G_j^3\,\bar G_j^2\right]_0\\
\label{eq_c24}
C_6:&\ &j=k=l=m\quad\to\quad
\frac{D^4\Gamma_1^2}{4\pi^2}\sum_j \left[G_j^3\,\bar G_j^3\right]_0
\sim \frac{1}{D}\,\frac{\Gamma_1^2}{(\Gamma_1-ir)^5}
\sim {\cal O}\left(\Pi_D^{-3}\right)\ .
\end{eqnarray}
We observe that (\ref{eq_c14}), (\ref{eq_c16}) and (\ref{eq_c19}) 
mutually cancel. Furthermore, (\ref{eq_c17}) is cancelled by 
(\ref{eq_c22}) and (\ref{eq_c23}). (\ref{eq_c24}) is already 
of third order and has to be disregarded. The only second order 
contributions are (\ref{eq_c20}) and (\ref{eq_c21}) which give
by (\ref{eq_c6}--\ref{eq_c8})
\begin{equation}
\label{eq_c25}
[B_1\,\bar B_1]_0=\frac{D^4\Gamma_1^2}{4\pi^2}\,\frac{1}{D^2}\,
\frac{4\pi^2}{(D\Gamma_1-i\omega)^4}\,
[2+i(-i)]=\frac{3}{D^2}\,\frac{\Gamma_1^2}{(\Gamma_1-ir)^4}\ .
\end{equation}
This expression and (\ref{eq_c12}) are just the extra terms mentioned 
in (\ref{eq61}).

\newpage

\begin{center}
{\bf \large Figures}
\end{center}

\end{document}